\documentclass{aa}
\usepackage[varg]{txfonts}

\usepackage{placeins}
\usepackage[colorlinks=true, linkcolor=blue, urlcolor=blue, citecolor=blue, anchorcolor=blue]{hyperref}

\usepackage{natbib}
\bibpunct{(}{)}{;}{a}{}{,}

\usepackage{orcidlink}

\title{The ESO SupJup Survey VII: Clouds and line asymmetries in CRIRES$^+$ J-band spectra of the Luhman 16 binary}
\titlerunning{J-band characterisation of Luhman 16 with CRIRES$^+$}

\author{
S. de Regt\inst{\ref{instLeiden}}\orcidlink{0000-0003-4760-6168} \and
I. A. G. Snellen\inst{\ref{instLeiden}}\orcidlink{0000-0003-1624-3667} \and
N. F. Allard\inst{\ref{instLIRA},\ref{instIAP}}\orcidlink{0000-0001-6220-6221} \and
D. Gonz\'alez Picos\inst{\ref{instLeiden}}\orcidlink{0000-0001-9282-9462} \and
S. Gandhi\inst{\ref{instWarwick},\ref{instCEH}}\orcidlink{0000-0001-9552-3709} \and
N. Grasser\inst{\ref{instLeiden}}\orcidlink{0009-0009-6634-1741} \and
R. Landman\inst{\ref{instLeiden}}\orcidlink{0000-0002-7261-8083} \and
P. Molli\`ere\inst{\ref{instMPIA}}\orcidlink{0000-0003-4096-7067} \and
E. Nasedkin\inst{\ref{instTCD}}\orcidlink{0000-0002-9792-3121} \and
T. Stolker\inst{\ref{instLeiden}}\orcidlink{0000-0002-5823-3072} \and
Y. Zhang\inst{\ref{instCalTech}}\orcidlink{0000-0003-0097-4414}
}

\institute{
Leiden Observatory, Leiden University, P.O. Box 9513, 2300 RA, Leiden, The Netherlands\\\email{regt@strw.leidenuniv.nl} \label{instLeiden} \and
LIRA, Observatoire de Paris, Universit\'e PSL, Sorbonne Universit\'e, Sorbonne Paris Cit\'e, CNRS, 61, Avenue de l'Observatoire, F-75014 Paris, France \label{instLIRA} \and
Institut d’Astrophysique de Paris, UMR7095, CNRS, Universit\'e Paris VI, 98bis Boulevard Arago, F-75014 Paris, France \label{instIAP} \and
Department of Physics, University of Warwick, Coventry CV4 7AL, UK \label{instWarwick} \and
Centre for Exoplanets and Habitability, University of Warwick, Gibbet Hill Road, Coventry CV4 7AL, UK \label{instCEH} \and
Max-Planck-Institut für Astronomie, Königstuhl 17, 69117 Heidelberg, Germany \label{instMPIA} \and
School of Physics, Trinity College Dublin, University of Dublin, Dublin 2, Ireland \label{instTCD} \and
Department of Astronomy, California Institute of Technology, Pasadena, CA 91125, USA \label{instCalTech}
}

\date{Received date / Accepted date}

\abstract
{Brown dwarfs at the L-T transition likely experience an inhomogeneous clearing of the clouds in their atmospheres. The resulting surface of thin and thick cloudy patches has been put forward to explain the observed variability, J-band brightening, and re-emergence of FeH absorption.} 
{We study the closest brown dwarf binary, Luhman 16A and B, in an effort to constrain their chemical and cloud compositions. As this binary consists of an L7.5 and T0.5 component, we gain insight into the atmospheric properties at the L-T transition.} 
{As part of the ESO SupJup Survey, we observed Luhman 16AB at high spectral resolution in the J-band ($1.1$--$1.4\ \mathrm{\mu m}$) using CRIRES$^+$. To analyse the spectra, we employ an atmospheric retrieval framework, coupling the radiative transfer code \texttt{petitRADTRANS} with the \texttt{MultiNest} sampling algorithm.}
{For both objects, we report detections of H$_2$O, K, Na, FeH, and, for the first time in the J-band, hydrogen-fluoride (HF). The K doublet at $1250\ \mathrm{nm}$ shows asymmetric absorption in the blue line wings, which are reproduced via pressure- and temperature-dependent shifts of the line cores. We find evidence for clouds in both spectra and we place constraints on an FeH-depletion in the Luhman 16A photosphere. The inferred over-abundance of FeH for Luhman 16B opposes its predicted rainout into iron clouds. A two-column model, which emulates the patchy surface expected at the L-T transition, is weakly preferred ($\sim$\,$1.8\sigma$) for component B but disfavoured for A ($\sim$\,$5.5\sigma$).}
{The results suggest a uniform surface on Luhman 16A, which is in good agreement with the reduced variability observed for this L-type component. While the presented evidence is not sufficient to draw conclusions about any inhomogeneity on Luhman 16B, future observations covering a broader wavelength range could help to test the cloud-clearing hypothesis.}


\usepackage{float}
\usepackage{subfiles}
\usepackage{pdflscape}
\usepackage{xcolor}
\definecolor{A1}{HTML}{FF622E}
\definecolor{B1}{HTML}{396ED8}

\definecolor{c1}{HTML}{000000}
\definecolor{c2}{HTML}{8C8C8C}

\usepackage{xurl}

\begin{document}

\maketitle

\section{Introduction}
As brown dwarfs cool down to effective temperatures below $\sim$\,$1300\ \mathrm{K}$, their spectro-photometric appearance evolves at the transition between L and T spectral types \citep{Kirkpatrick_ea_2005}. The dissipation of clouds in the photospheres of the cooler T-type objects can explain the observed blue-ward shift in colour \citep{Burrows_ea_2006,Saumon_ea_2008}. In addition, brown dwarfs at the L-T transition are often variable, pointing towards disruptions of the cloudy surface \citep{Radigan_ea_2014,Liu_ea_2024}. Further, the re-emergence of FeH absorption observed in early- to mid-T type spectra \citep{Nakajima_ea_2004,Cushing_ea_2005} suggests that cloudless or transparent patches are formed in their photospheres. These clearer regions are thought to expose deeper, hotter layers where FeH has not yet condensed into an iron cloud deck \citep{Burgasser_ea_2002,Lodders_ea_2006}. 

The opacity cross-sections of molecular gases, such as H$_2$O and CH$_4$, generally decrease towards shorter wavelengths (e.g. \citealt{Morley_ea_2014,Gandhi_ea_2020}). Consequently, observations in the J-band ($1.1$--$1.4\ \mathrm{\mu m}$) can probe condensate clouds that reside deep in the atmosphere ($\sim$\,$10\ \mathrm{bar}$; \citealt{Ackerman_ea_2001,Marley_ea_2002}). This increased sensitivity and the cloud-clearing expected for early- to mid-T dwarfs can explain why these spectral types exhibit a brightening of the J-band magnitudes \citep{Burgasser_ea_2002,Marley_ea_2010,Dupuy_ea_2012}. Similarly, the high J-band variability at the L-T transition follows from the expected cloud inhomogeneities \citep{Artigau_ea_2009,Radigan_ea_2012,McCarthy_ea_2024}. Furthermore, the high pressures probed can significantly broaden spectral lines, particularly for the alkali metals Na and K (e.g. \citealt{Schweitzer_ea_1996,Allard_ea_2007}), which have strong absorption lines in the J-band. The alkali line-widths can therefore serve as useful indicators of the surface gravity \citep{McLean_ea_2007,Allers_ea_2013}.

WISE J104915.57-531906.1, or Luhman 16, is the nearest known brown dwarf binary at a distance of $\sim$\,$2\ \mathrm{pc}$ \citep{Luhman_2013,Bedin_ea_2024}. The primary and secondary components coincide with the L-T transition, with an L7.5 and T0.5$\pm$1.0 classification, respectively \citep{Burgasser_ea_2013}. \citet{Bedin_ea_2024} determined dynamical masses of $M_\mathrm{A}=35.4\pm0.2$ and $M_\mathrm{B}=29.4\pm0.2\ M_\mathrm{Jup}$, consistent with the detection of lithium reported by \citet{Faherty_ea_2014} and \citet{Lodieu_ea_2015}. These low masses and the discovery of its membership to the Oceanus moving group, indicate an age of $\sim$\,$500\ \mathrm{Myr}$ \citep{Garcia_ea_2017,Gagne_ea_2023}. Despite being fainter in the K-band, Luhman 16B has a higher J-band flux than the primary component \citep{Burgasser_ea_2013}. Since the brown dwarfs share similar effective temperatures ($T_\mathrm{eff}\sim$\,$1200$--$1300\ \mathrm{K}$; \citealt{Faherty_ea_2014,Lodieu_ea_2015}), this flux-reversal hints at a lower cloud opacity for Luhman 16B \citep{Burgasser_ea_2013}. In keeping with the cloud-clearing hypothesis, both components are variable, but a stronger variability has been established for Luhman 16B (e.g. \citealt{Biller_ea_2013,Biller_ea_2024,Buenzli_ea_2015,Fuda_ea_2024}). Additionally, Doppler imaging analyses by \citet{Crossfield_ea_2014} and \citet{Chen_ea_2024} uncovered prominent surface inhomogeneities on Luhman 16B. 

In this work, we present an atmospheric retrieval analysis of Luhman 16AB, using high-resolution J-band spectra. Section \ref{sect:methods} outlines the reduction of our observations and the utilised modelling framework. In Sect. \ref{sect:results}, we describe the retrieved results and a discussion is provided in Sect. \ref{sect:discussion}. Section \ref{sect:conclusions} summarises the main conclusions drawn from our retrieval analysis.

\begin{figure*}[h!]
    \centering
    \includegraphics[width=17cm]{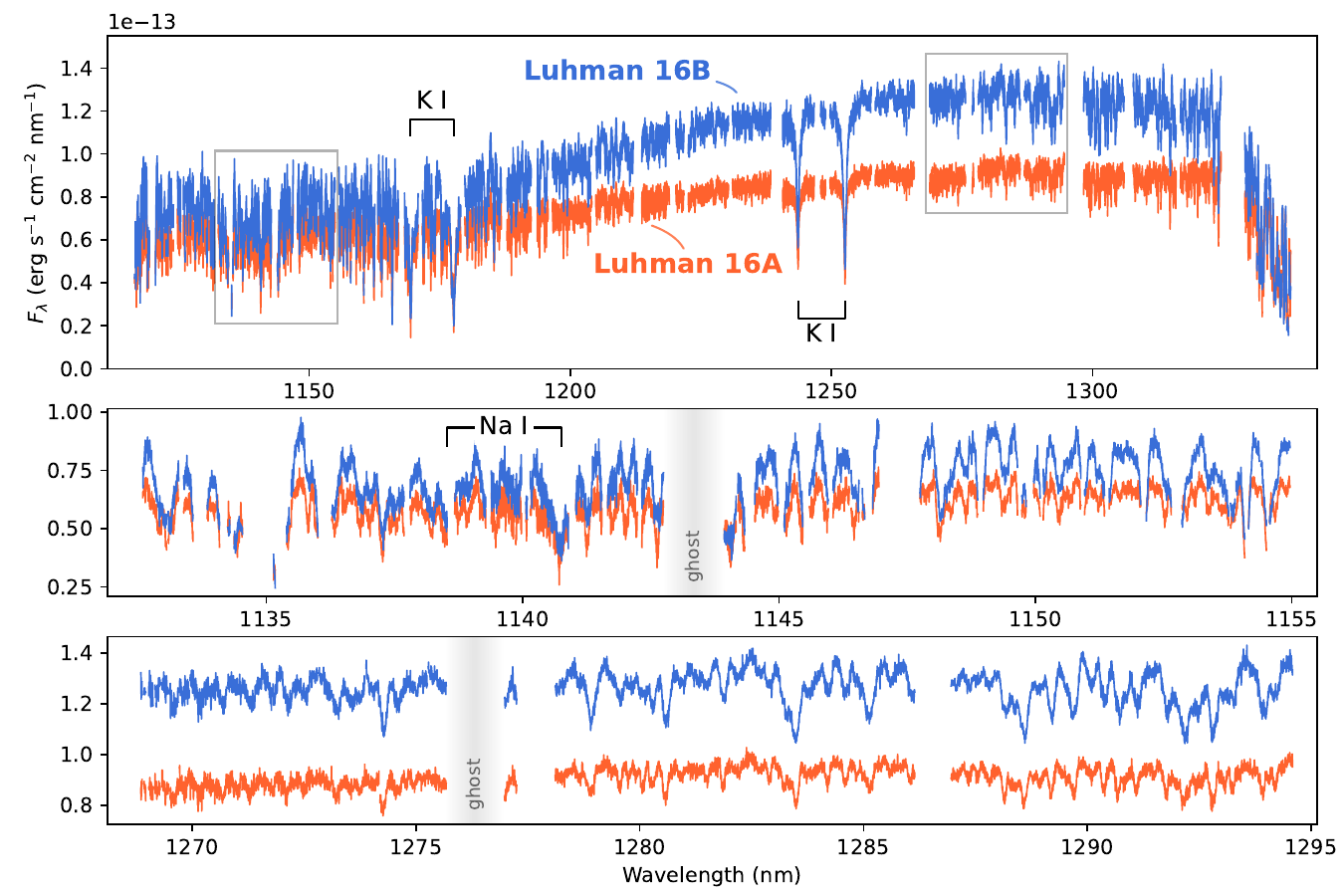}
    \caption{CRIRES$^+$ J-band spectra of the Luhman 16 binary. \textit{Top panel}: nine spectral orders covered in the J1226 wavelength setting. The two reddest order-detector pairs are omitted, as described in Sect. \ref{sect:obs_and_red}. \textit{Lower panels}: zoom-ins showing the similar absorption features between the brown dwarfs.}
    \label{fig:spectrum}
\end{figure*}

\section{Methods} \label{sect:methods}
\subsection{Observations and reduction} \label{sect:obs_and_red}
The Luhman 16 binary was observed with the CRyogenic high-resolution InfraRed Echelle Spectrograph (CRIRES$^+$; \citealt{Dorn_ea_2023}) on January 1, 2023, as part of the ESO SupJup Survey (Program ID: 110.23RW; see \citealt{de_Regt_ea_2024}). Only a single nodding cycle was performed in the J1226 wavelength setting, resulting in 4 exposures of $300\ \mathrm{s}$ each. The $0.4"$ slit was used to obtain signal-to-noises (at $\sim$\,$1285\ \mathrm{nm}$) of $90$ and $110$ for Luhman 16A and B, respectively. We refer to \citetext{de Regt et al., in prep.} for a detailed analysis of our CRIRES$^+$ K-band observations of Luhman 16. Following \citet{de_Regt_ea_2024}, the spectra were reduced with \texttt{excalibuhr}\footnote{\url{https://github.com/yapenzhang/excalibuhr}} \citep{Zhang_ea_2024}. Subsequent to the dark-subtraction, flat-fielding and sky-subtraction, we extracted the spectra of the binary components separately using 12-pixel wide apertures. At the time of observation, the components had a projected separation of $\sim$\,$0.81"$ \citep{Bedin_ea_2024}. Combined with the optical seeing of $\sim$\,$0.9"$ and the absence of adaptive-optics assistance (limited by the R-band magnitude), this resulted in a degree of blending between the spectral traces as demonstrated in Fig. \ref{fig:blend_corr}. To mitigate this, we applied a correction for the contamination in the extraction apertures by fitting and integrating the spatial profiles, as is outlined in Appendix \ref{app:blend_corr}.

The standard star HD 93563 was observed after Luhman 16 using the same observing setup. Its extracted spectrum provides a high-quality measurement of telluric absorption lines and the instrumental throughput. Similar to \citet{Gonzalez_Picos_ea_2024}, we applied \texttt{molecfit} \citep{Smette_ea_2015} to the standard spectrum and used the resulting model to correct the telluric absorption lines in the Luhman 16AB spectra. As the deepest tellurics are poorly fit, we masked out all pixels with a transmissivity below $\mathcal{T}<70\%$. While fitting the tellurics, \texttt{molecfit} estimates the continuum using first-degree polynomial fits. We recovered the instrumental throughput by dividing this continuum with the blackbody spectrum associated with the standard star ($T_\mathrm{eff}=15\,000\ \mathrm{K}$; \citealt{Arcos_ea_2018}). We subsequently corrected for the wavelength-dependent throughput in the Luhman 16 spectra. \texttt{Molecfit} also provides an estimated resolution of $R=\lambda/\Delta\lambda\sim65\,000$ ($\sim$\,$4.6\ \mathrm{km\ s^{-1}}$) and an optimised wavelength solution, which we adopted for the target spectra.

For plotting purposes, we performed a flux-calibration by scaling the Luhman 16 spectra to match the broadband photometry ($\mathrm{J_A}=11.53\pm0.04$ and $\mathrm{J_B}=11.22\pm0.04\ \mathrm{mag}$; \citealt{Burgasser_ea_2013}). The reduced J-band spectra are shown in Fig. \ref{fig:spectrum} in orange and blue for Luhman 16A and B, respectively. In the J1226 setting, the traces of the reddest spectral order are projected (partially) outside of the detectors and thus we omit the last two detectors from our analysis, leaving $8\cdot3+1=25$ usable order-detector pairs. Furthermore, J1226 observations contain a ghost in the bluest detector which we mask out using the wavelength ranges reported in the CRIRES$^+$ manual\footnote{\url{https://www.eso.org/sci/facilities/paranal/instruments/crires/doc/ESO-254264_CRIRES_User_Manual_P110.2.pdf}} and an additional $0.1\ \mathrm{nm}$ spacing on either side. The spectra in Fig. \ref{fig:spectrum} exhibit notable features, such as the potassium doublets at $\sim$\,$1175\ \mathrm{nm}$ and $\sim$\,$1250\ \mathrm{nm}$. The comparable spectral features in the zoomed-in panels underline the similar spectral types across the binary, while a difference in line-broadening reveals distinct (projected) rotational velocities.

\subsection{Retrieval framework}
To infer the atmospheric properties of the Luhman 16 binary, we used a retrieval framework where the radiative transfer code \texttt{petitRADTRANS} (\texttt{pRT}; version 3.1; \citealt{Molliere_ea_2019,Molliere_ea_2020,Alei_ea_2022}) is coupled to the \texttt{PyMultiNest} nested sampling code \citep{Feroz_ea_2009,Buchner_ea_2014}. The retrievals utilise 1000 live points at a constant sampling efficiency of $5\%$. The retrieved parameters and their priors are listed in Table \ref{tab:params}, some of which are discussed below in more detail. 

\subsubsection{Likelihood and covariance} \label{sect:likelihood}
Following \citet{Ruffio_ea_2019}, we defined the likelihood as
\begin{align}
    \ln{\mathcal{L}} &= \sum_i \ln{\mathcal{L}_i} \\
    \ln{\mathcal{L}_i} &= -\frac{1}{2}\left(N_i\ln\left(2\pi s_i^2\right)+\ln\left(|\vec{\Sigma}_{0,i}|\right)+\frac{1}{s_i^2}\vec{r}_i^T\vec{\Sigma}_{0,i}^{-1}\vec{r}_i\right), 
\end{align}
where we sum over the 25 order-detector pairs $i$ with $N_i$ valid pixels ($2048$ at most). For each chip, $\vec{\Sigma}_{0,i}$ is the un-scaled covariance matrix and $\vec{r}_i$ denotes the residuals between the data $\vec{d}_i$ and model $\vec{m}_i$, calculated as
\begin{align}
    \vec{r}_i &= \vec{d}_i - \phi_i \vec{m}_i 
\end{align}
The optimal flux- and covariance-scaling parameters, $\tilde{\phi_i}$ and $\tilde{s_i}^2$, are calculated at each likelihood evaluation via
\begin{align}
    \tilde{\phi_i} &= \left(\vec{m}_i^T\vec{\Sigma}_{0,i}^{-1}\vec{m}_i\right)^{-1} \vec{m}_i^T\vec{\Sigma}_{0,i}^{-1}\vec{d}_i, \\
    \tilde{s_i^2} &= \frac{1}{N_i} \left(\vec{d}_i - \tilde{\phi_i} \vec{m}_i\right)^T\vec{\Sigma}_{0,i}^{-1} \left(\vec{d}_i - \tilde{\phi_i} \vec{m}_i\right).
\end{align}

Following \citet{de_Regt_ea_2024}, we employ Gaussian Processes (GP) to model correlated noise between neighbouring pixels. We add a radial basis function to the covariance matrix as
\begin{align}
    \vec{\Sigma}_{0,i} &= \vec{\sigma}_i^2\vec{I} + a^2 \sigma_\mathrm{eff,i}^2 \exp{\left(-\frac{\left(\vec{\lambda}_i^T-\vec{\lambda}_i\right)^2}{2\ell^2}\right)}, \\
    \sigma_\mathrm{eff,i} &= \mathrm{median}(\vec{\sigma}_i), 
\end{align}
where $\vec{\sigma}_i$ is the flux-uncertainty vector, and $\vec{\lambda}_i^T-\vec{\lambda}_i$ results in a distance matrix between all pairs of pixels (in $\mathrm{nm}$). The GP amplitude $a$ and length-scale $\ell$, which modulates the correlation distance, are retrieved as free parameters.

\subsubsection{Surface gravity}
The orbit of Luhman 16AB has been extensively studied (e.g. \citealt{Bedin_ea_2017, Garcia_ea_2017, Lazorenko_ea_2018, Bedin_ea_2024}), resulting in well-constrained dynamical masses for both components. By imposing this information as prior constraints, we can break the degeneracy that retrieval studies frequently find between the surface gravity and metallicity (e.g. \citealt{Zhang_ea_2021,Zhang_ea_2023,de_Regt_ea_2024, Gonzalez_Picos_ea_2024}). The surface gravity is related to the brown dwarf mass, $M$, and radius, $R$, via 
\begin{align}
    \textit{g} = \frac{GM}{R^2},
\end{align}
with $G$ the gravitational constant. We adopt masses of $M_\mathrm{A}=35.4\pm0.2$ and $M_\mathrm{B}=29.4\pm0.2\ M_\mathrm{Jup}$ \citep{Bedin_ea_2024} and a radius of $R=1\pm0.1\ R_\mathrm{Jup}$ \citep{Biller_ea_2024}. In combination, this results in Gaussian priors of $\log\textit{g}_\mathrm{A}=4.96\pm0.09$ and $\log\textit{g}_\mathrm{B}=4.88\pm0.09$ on the surface gravity.

\subsubsection{Pressure-temperature profile}
Akin to \citet{Zhang_ea_2023}, we model the thermal profile by parameterising the temperature gradient $\nabla_i=\frac{d\ln{T_i}}{d\ln{P_i}}$ at five points in log-pressure space. The outer knots are set to the limits of the modelled atmospheric layers ($P_1=10^3$ and $P_5=10^{-5}\ \mathrm{bar}$), while the three intermediate knots are allowed to shift to avoid biasing the photosphere towards certain pressures \citep{Gonzalez_Picos_ea_2025}. The temperature gradients $\nabla_j$ at each of the 50 atmospheric layers in our \texttt{pRT} model are obtained through linear interpolation from the values at the knots $\nabla_i$. Subsequently, the temperature at each of the layers $j$ is calculated via
\begin{align}
    T_{j} &= T_{j-1}\cdot\left(\frac{P_{j}}{P_{j-1}}\right)^{\nabla_{j}}. 
\end{align}
This parameterisation requires an anchor point for the temperature, which we determine at the central knot ($i=3$). As such, we retrieve nine parameters to describe the thermal profile: five gradients $\nabla_{i=1,2,3,4,5}$, three pressure points $P_{i=2,3,4}$, and one temperature $T_{i=3}$. This gradient-based parameterisation is flexible but also allows for the incorporation of radiative-convective equilibrium through the chosen prior constraints, thereby avoiding the excessively steep gradients or inversions that can arise when directly retrieving the temperature (e.g. \citealt{Line_ea_2015,Molliere_ea_2020,de_Regt_ea_2024}). 

\subsubsection{Clouds}
The pressures and temperatures probed with our J-band spectra cover the condensation curves of several cloud species (e.g. Fe, Mg$_2$SiO$_4$, MgSiO$_3$; \citealt{Visscher_ea_2010}). Hence, we model the total opacity of a grey cloud as
\begin{align}
    \kappa_\mathrm{cl}(P) &= \begin{cases}
        \kappa_\mathrm{cl,0} \cdot \left(\dfrac{P}{P_\mathrm{cl,0}}\right)^{f_\mathrm{sed}} & P<P_\mathrm{cl,0}, \\
        0 & P\geq P_\mathrm{cl,0}. \\
    \end{cases}
\end{align}
Here, $\kappa_\mathrm{cl,0}$ is the opacity at the cloud base, set by $P_\mathrm{cl,0}$, and $f_\mathrm{sed}$ governs the opacity decay above this base \citep{Molliere_ea_2020}. This simple parameterisation limits the number of free parameters and avoids assumptions about the chemical or morphological composition of cloud particles. Although defined as opacity rather than optical depth, this cloud parameterisation is comparable to the slab clouds used in the retrieval frameworks of \citet{Burningham_ea_2017} and \citet{Vos_ea_2023}. We leave out any wavelength-dependence as well as additional cloud layers due to the narrow wavelength extent of the J-band ($\sim$\,$0.2\ \mathrm{\mu m}$). Test retrievals where $\kappa_\mathrm{cl}(P)$ was divided into a scattering and absorption component did not yield substantially different results from including the cloud only as absorption. The single-scattering albedo used in these tests, $\omega$, also remained unconstrained. For that reason, we include $\kappa_\mathrm{cl}(P)$ only as an absorption opacity in this work.

\subsubsection{Chemistry}
We model the chemical abundances using a free-chemistry approach, where we retrieve a volume-mixing ratio (VMR) for each of the included chemical species: H$_2$O, FeH, HF, and the alkalis Na and K. The abundance is held constant vertically, except for FeH, which is expected to condense into iron cloud particles around the probed pressures (e.g. \citealt{Visscher_ea_2010}). Instead, we mimic the rapidly decreasing FeH abundance via
\begin{align}
    \mathrm{VMR(FeH)}(P) = \begin{cases}
            \mathrm{VMR(FeH)_0}\left(\dfrac{P}{P_{\mathrm{FeH},0}}\right)^{\alpha_\mathrm{FeH}} & P<P_{\mathrm{FeH},0}, \\
            \mathrm{VMR(FeH)_0} & P\geq P_{\mathrm{FeH},0}, \\
        \end{cases}
\end{align}
where $\mathrm{VMR(FeH)}_0$ is the FeH abundance of the lower atmosphere, $P_{\mathrm{FeH},0}$ is the pressure above which the abundance falls of with a power-law, determined by $\alpha_\mathrm{FeH}$. Sodium (Na) and potassium (K) should also form clouds (Na$_2$S(s,l), KCl(s,l), KAlSi$_3$O$_8$(s); \citealt{Kitzmann_ea_2024}), but at higher altitudes so we decided to not adopt the same rainout treatment to minimise the number of free parameters. The abundance of He is set to $\mathrm{VMR(He)}=0.15$ and that of H$_2$ is adjusted to ensure a total mixing ratio equal to unity. Other species (e.g. CrH, TiO, VO, CH$_4$, NH$_3$, Fe) were included in test retrievals but did not yield reasonable constraints on the abundances and we therefore omit them from the presented analysis. 

\subsection{Opacity cross-sections}
Our \texttt{pRT} model accounts for several sources of opacity, including collision-induced absorption from H$_2$-H$_2$ and H$_2$-He pairs as well as Rayleigh scattering from H$_2$ and He, the most abundant chemical species in a sub-stellar atmosphere. For the molecular opacities of H$_2$O \citep{Polyansky_ea_2018}, FeH \citep{Dulick_ea_2003,Bernath_2020}, and HF \citep{Li_ea_2013,Coxon_ea_2015,Somogyi_ea_2021}, we use tables that are pre-computed on a grid of pressures and temperatures\footnote{Using line lists obtained from the ExoMol database: \url{https://www.exomol.com/}}. The perturbations brought on by collisions with H$_2$ and He are most apparent for the strong Na and K lines. As such, we employ a more detailed treatment for the alkali opacity cross-sections. 

\subsubsection{Parameterised widths and shifts for K $\mathrm{4p-5s}$ doublet} \label{sect:width_shift_param}
In addition to broadening the line profiles, collisions with H$_2$ and He perturb the energy potential of the radiator thus resulting in wavelength-shifts of the line centres. The impact theories of pressure broadening \citep{Baranger_ea_1958,Kolb_ea_1958} are based on the assumption of sudden collisions between the radiator and perturbing atoms, and are valid when frequency displacements and gas densities are sufficiently small. In impact broadening, the duration of the collision is assumed to be small compared to the interval between collisions, and the results describe the line within a few line widths of the centre. 

We use the unified line shape theory of \citet{Allard_ea_1999}, incorporating more accurate potentials than van der Waals approximations, to determine the line parameters of the cores of the $1250\ \mathrm{nm}$ K doublet, perturbed by H$_2$ and He. These lines are formed from the $\mathrm{4p ^2P_{3/2}-5s}$ ($\nu_0=7983.7\ \mathrm{cm}^{-1}$) and $\mathrm{4p ^2P_{1/2}-5s}$ ($\nu_0=8041.4\ \mathrm{cm^{-1}}$) transitions. For K--He, the molecular-structure calculations performed by \citet{Pascale_ea_1983} are used for the adiabatic potential of the 5s state and we use a combination of ab initio potentials of \citet{Santra_ea_2005} and \citet{Nakayama_ea_2001} for the 4p state (see \citealt{Allard_ea_2024}). Molecular data for the 4p and 5s states of K--H$_2$ were described in \citet{Allard_ea_2016}. An average over velocity was done numerically by performing the calculation for different velocities and then thermally averaging with 24-point Gauss-Laguerre integration. This impact approximation can be used to determine the line core for H$_2$ or He densities below $n_p=10^{19}\ \mathrm{cm^{-3}}$.

\begin{figure}[h!]
    \centering
    \resizebox{\hsize}{!}{\includegraphics[width=17cm]{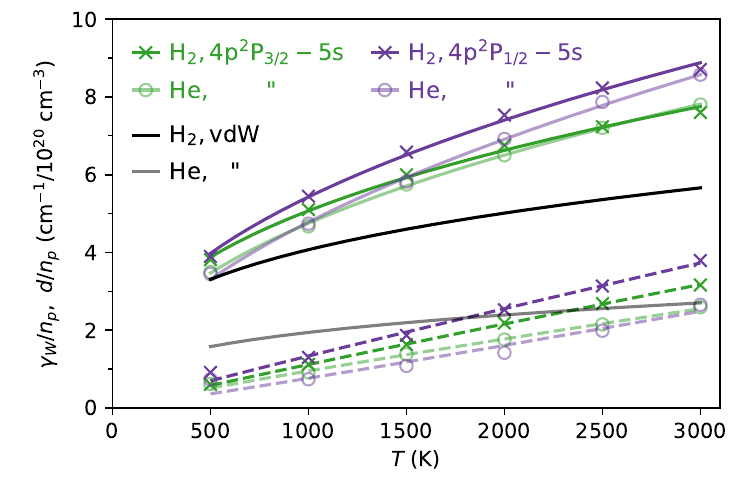}}
    \caption{Variation with temperature of the broadening ($\gamma_W/n_p$, solid) and shift ($d/n_p$, dashed) rates of the $\mathrm{4p^2P_{3/2}-5s}$ (green) and $\mathrm{4p^2P_{1/2}-5s}$ (purple) lines, perturbed by H$_2$ and He. The van der Waals (vdW) broadening, computed with Eq. \ref{eq:width_Kurucz}, is shown in black for comparison.}
    \label{fig:core_power_law}
\end{figure}

\begin{table}[h!]
    \centering
    \caption{Broadening- and shift-parameters used in Eqs. \ref{eq:width} and \ref{eq:shift} for the potassium doublet at $1250\ \mathrm{nm}$.}
    \label{tab:broad_params}
    {\small
    \begin{tabular}{r|cc|cc}
        \hline\hline
         & \multicolumn{2}{c|}{$\mathrm{4p^2P_{3/2}-5s}$} & \multicolumn{2}{c}{$\mathrm{4p^2P_{1/2}-5s}$} \\
         & H$_2$ & He & H$_2$ & He \\
        \hline
        $A_{W}$ & 0.352609 & 0.20819 & 0.245926 & 0.121448 \\
        $b_{W}$ & 0.385961 & 0.452833 & 0.447971 & 0.531718 \\
        $A_{d}$ & 0.00158988 & 0.00194382 & 0.00211668 & 0.000462539 \\
        $b_{d}$ & 0.949254 & 0.89691 & 0.933563 & 1.07284 \\
    \end{tabular}
    }
\end{table}

\begin{figure}[h!]
    \centering
    \resizebox{\hsize}{!}{\includegraphics[width=17cm]{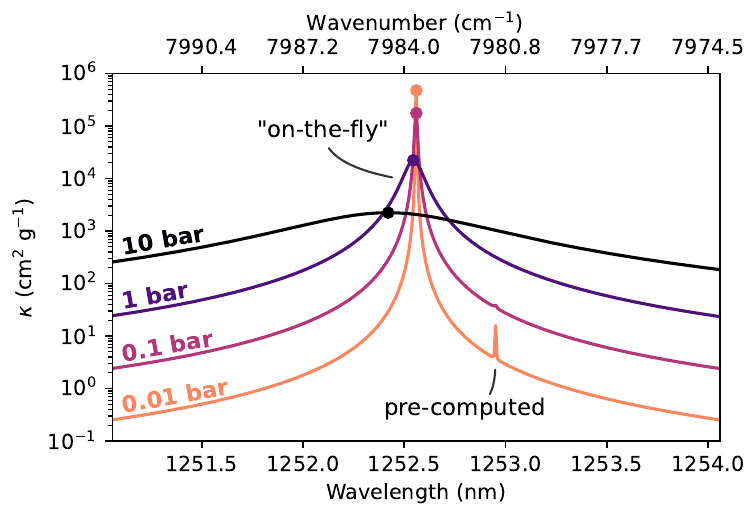}}
    \caption{Opacity cross-sections for potassium (K) at $T=1500\ \mathrm{K}$. The weak line at $1252.95\ \mathrm{nm}$ is interpolated from a pre-computed opacity table. The doublet line at $1252.55\ \mathrm{nm}$ ($\mathrm{4p^2P_{3/2}-5s}$) is computed "on-the-fly", meaning that the line-strength and -width are calculated exactly for each $PT$-point in the modelled atmosphere. This line also receives a $PT$-dependent shift, resulting in a blue-shift at high pressures.}
    \label{fig:line_shift}
\end{figure}

The determined line-widths and shifts are presented in Fig. \ref{fig:core_power_law} as crosses and circles for a range of temperatures, for both lines, and for H$_2$ and He perturbers. The line width, $\gamma_W$, is measured by half the full width at half the maximum intensity, what is customarily termed HWHM. The solid and dashed lines in Fig. \ref{fig:core_power_law} show the power-law fits, defined as 
\begin{align}
    \gamma_W[\mathrm{cm^{-1}}] &= \sum_{p\in\{\mathrm{H_2,He}\}} A_{W,p} T^{b_{W,p}} \left(\frac{n_p}{10^{20}\ \mathrm{cm^{-3}}}\right), \label{eq:width} \\
    d[\mathrm{cm^{-1}}] &= \sum_{p\in\{\mathrm{H_2,He}\}} A_{d,p} T^{b_{d,p}} \left(\frac{n_p}{10^{20}\ \mathrm{cm^{-3}}}\right). \label{eq:shift}
\end{align}
Here, $d$ is the wavenumber-shift to be added to the line centre $\nu_0$, and $n_p$ is the number density of the perturber \citep{Allard_ea_2023}. The widths and shifts are thus linearly dependent on H$_2$/He densities, and a power law in temperature described by the $A_{W/d}$ and $b_{W/d}$ coefficients. The perturber-specific coefficients for both transitions are given in Table \ref{tab:broad_params}. Utilising these parameters results in blue-shifted line profiles with increasing pressures, as is demonstrated for the $\mathrm{4p^2P_{3/2}-5s}$ line in Fig. \ref{fig:line_shift}.

\subsubsection{"On-the-fly" line profiles} \label{sect:on_the_fly}
During the retrieval, \texttt{pRT} interpolates the pre-computed molecular cross-section of H$_2$O, FeH, and HF onto the pressures and temperatures of the model atmospheres. This conventional interpolation method bypasses calculating thousands of line profiles at each of the atmospheric layers, thereby drastically reducing the computation time. However, we found that the nominal pressure-spacing of $1\ \mathrm{dex}$ is too sparse to properly fit the strongly pressure-broadened Na and K doublets (see Sect. \ref{sect:alkali_discussion}). Since there are few strong alkali lines in the J-band wavelength range, it is computationally feasible to use a more exact method to obtain their opacities. 

We acquired energy, oscillator-strength, and broadening information from the Kurucz\footnote{\url{http://kurucz.harvard.edu/atoms.html}} and NIST databases\footnote{\url{https://physics.nist.gov/PhysRefData/ASD/levels_form.html}} for neutral Na and K \citep{Kurucz_ea_2018}. During the retrieval, we interpolate the opacity from weak lines (with oscillator strengths $\log \textit{gf}<-0.5$) from a pre-computed table. Stronger lines are calculated "on-the-fly" at the temperature and pressure of each atmospheric layer, resulting in accurate thermal- and pressure-broadening prescriptions. We implement these opacities using the \texttt{additional\_absorption\_opacities\_function} functionality in \texttt{pRT}. Following \citet{Sharp_ea_2007} and \citet{Gandhi_ea_2020}, we define the Gaussian and Lorentzian half-widths as
\begin{align}
    \gamma_G[\mathrm{cm^{-1}}] &= \frac{\nu_0}{c} \sqrt{\frac{2 k_B T}{m_{X}}},  \\ 
    \gamma_L[\mathrm{cm^{-1}}] &= \gamma_N + \sum_{p\in\{\mathrm{H_2,He}\}}\gamma_{W,p}, 
\end{align}
where $\nu_0$ is the transition energy, $m_{X}$ the atomic mass of Na or K (in $\mathrm{g}$), and $\gamma_N$ is the natural broadening which is negligible at the high pressures of the Luhman 16 photospheres. For all but the K $\mathrm{4p-5s}$ lines, we calculate the half-width per perturber following Eq. 23 in \citet{Sharp_ea_2007}.
\begin{align}
    \gamma_{W,p}[\mathrm{cm^{-1}}] &= \frac{\gamma_\mathrm{vdW}^\mathrm{Kurucz}}{4\pi c} \left(\frac{\mu_{\mathrm{H},X}}{\mu_{p,X}}\frac{T}{10\,000\ \mathrm{K}}\right)^{3/10} \left(\frac{\alpha_p}{\alpha_\mathrm{H}}\right)^{2/5}\cdot n_p, \label{eq:width_Kurucz}
 \end{align}
where $\gamma_\mathrm{vdW}^\mathrm{Kurucz}$ is the van der Waals coefficient provided in the Kurucz line list, $\mu_{a,b}=m_a m_b / (m_a+m_b)$ is the reduced mass (in $\mathrm{g}$), and $\alpha_a$ is the polarisability of species $a$ (in $\mathrm{cm^3}$). The black (H$_2$) and grey (He) lines in Fig. \ref{fig:core_power_law} show the broadening rates calculated with Eq. \ref{eq:width_Kurucz} for the K $\mathrm{4p-5s}$ transitions. The Kurucz line list reports the same $\gamma_\mathrm{vdW}^\mathrm{Kurucz}$\,$=10^{-7.46}\ \mathrm{s^{-1}\ cm^{3}}$ for both lines. Figure \ref{fig:core_power_law} reveals that the van der Waals approximation would underpredict the pressure broadening compared to the method outlined in Sect. \ref{sect:width_shift_param}.

\begin{figure*}[h!]
    \centering
    \includegraphics[width=17cm]{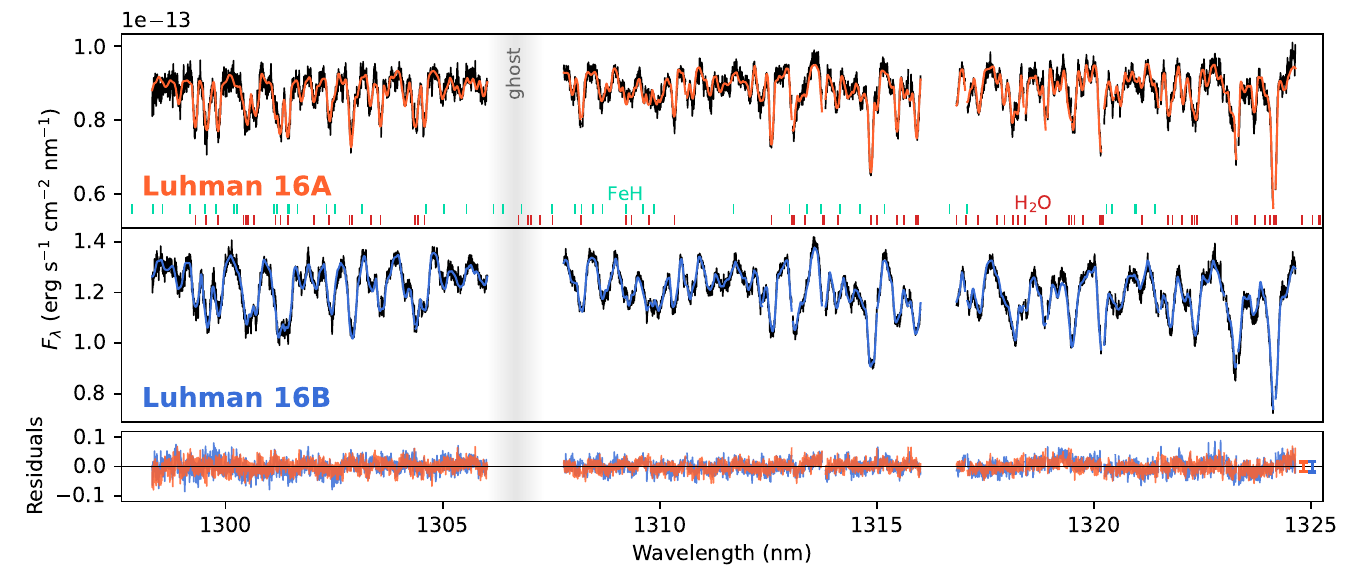}
    \caption{Best-fitting models to the J-band spectra of Luhman 16. The observations are shown in black, while the fiducial models are shown in orange and blue for Luhman 16A and B, respectively. The mean scaled uncertainties, defined as $\mathrm{diag}\left(\tilde{s_i}\sqrt{\vec{\Sigma}_{0,i}}\right)$, are displayed to the right of the residuals for reference. The predominant absorption lines of FeH and H$_2$O are indicated in the upper panel. The fits to the other spectral orders can be found in Appendix \ref{app:best_fitting_spectra}.}
    \label{fig:bestfit_spectrum}
\end{figure*}

\section{Results} \label{sect:results}
Our fiducial model provides a good fit to the Luhman 16 binary spectra, as presented for one spectral order in Fig. \ref{fig:bestfit_spectrum}. The full wavelength coverage is found in Appendix \ref{app:best_fitting_spectra}. The residuals mostly fall within the expected uncertainties (shown as errorbars), which are inflated during the fitting by a factor of $\tilde{s}^2\sim1.2$--$1.5$ to obtain a reduced $\chi^2$ equal to unity (see e.g. \citealt{Ruffio_ea_2019,Gibson_ea_2020}). We suspect that the minor model deficiencies (e.g. $\sim$\,$1192$--$1196$, $1211$, $1281$, $1292$, $1319$--$1324$, and $1334\ \mathrm{nm}$) arise due to missing absorbers or inaccuracies in the utilised line lists. Including other molecular or atomic species (e.g. CrH, TiO, VO, CH$_4$, NH$_3$, Fe), however, did not improve the fit or produce reasonable constraints on the abundances. 

\begin{figure*}[h!]
    \centering
    \includegraphics[width=17cm]{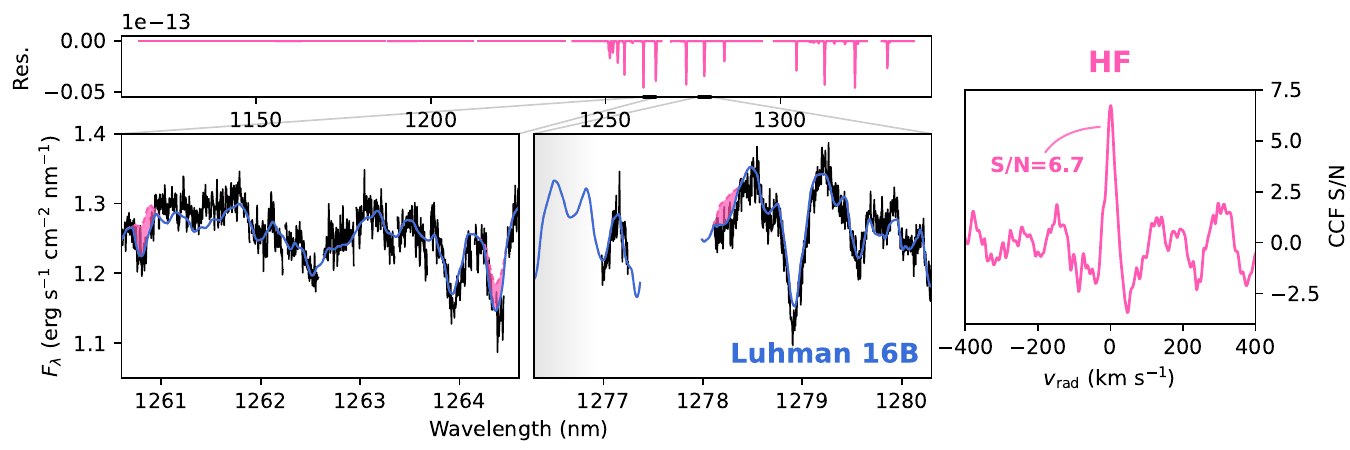}
    \caption{Detection analysis of hydrogen-fluoride (HF) in the Luhman 16B spectrum. The left panels show the spectral contribution of HF by comparing the complete model (blue, solid line) to a model without HF (pink, dashed line). The residuals display the difference between these two models. The right panel presents the cross-correlation function described in Sect. \ref{sect:res_CCF}.}
    \label{fig:HF_detection}
\end{figure*}

\subsection{Detection of chemical species}\label{sect:res_CCF}
From a visual inspection, we identify lines from H$_2$O, K and Na in the spectra of both brown dwarfs. As noted by \citet{Faherty_ea_2014} and \citet{Lodieu_ea_2015}, the Na doublet at $\sim$\,$1140\ \mathrm{nm}$ is weak and contaminated by (telluric) H$_2$O absorption, but is definitely present in our CRIRES$^+$ spectra (see Fig. \ref{fig:spectrum}). The $1250\ \mathrm{nm}$ K doublet of Luhman 16B shows stronger absorption compared to the primary, in line with previous studies covering the J-band \citep{Burgasser_ea_2013,Faherty_ea_2014,Lodieu_ea_2015}. Contrary to \citet{Faherty_ea_2014} and \citet{Lodieu_ea_2015}, we do not detect CH$_4$ with the up-to-date ExoMol line list \citep{Yurchenko_ea_2024}. This discrepancy might result from a fluctuation in the disequilibrium chemistry, where CH$_4$ would play a role (e.g. \citealt{Zahnle_ea_2014,Lee_ea_2024}), but a comparison with the X-SHOOTER data of \citet{Lodieu_ea_2015} revealed no major spectral differences. 

\begin{figure*}[h!]
    \centering
    \includegraphics[width=17cm]{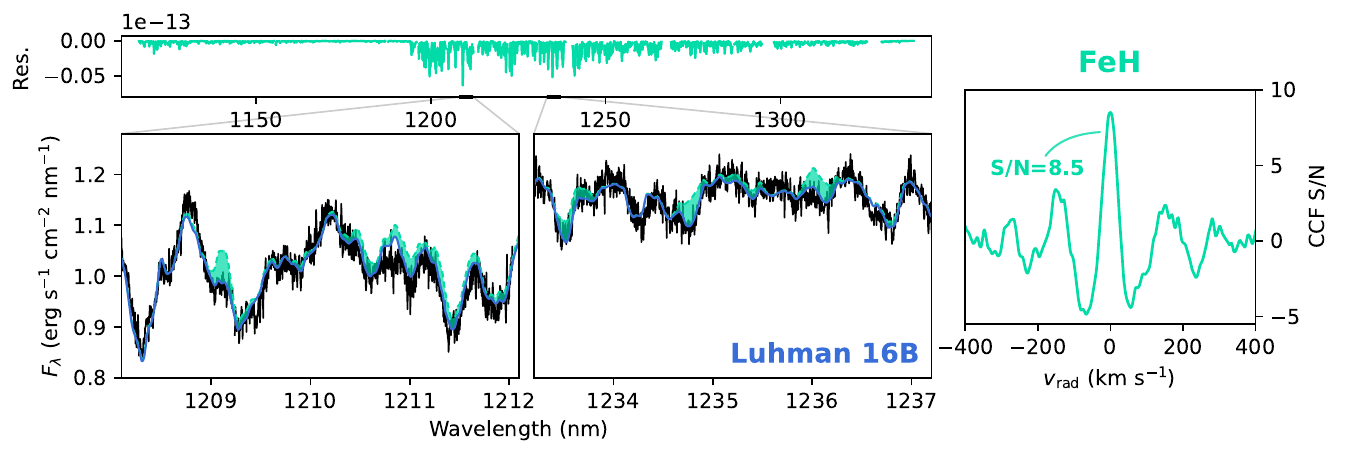}
    \caption{Same as Fig. \ref{fig:HF_detection}, but for iron-hydride (FeH). The model spectrum without FeH is shown as the turquoise, dashed line.}
    \label{fig:FeH_detection}
\end{figure*}

\begin{figure*}[h!]
    \centering
    \includegraphics[width=17cm]{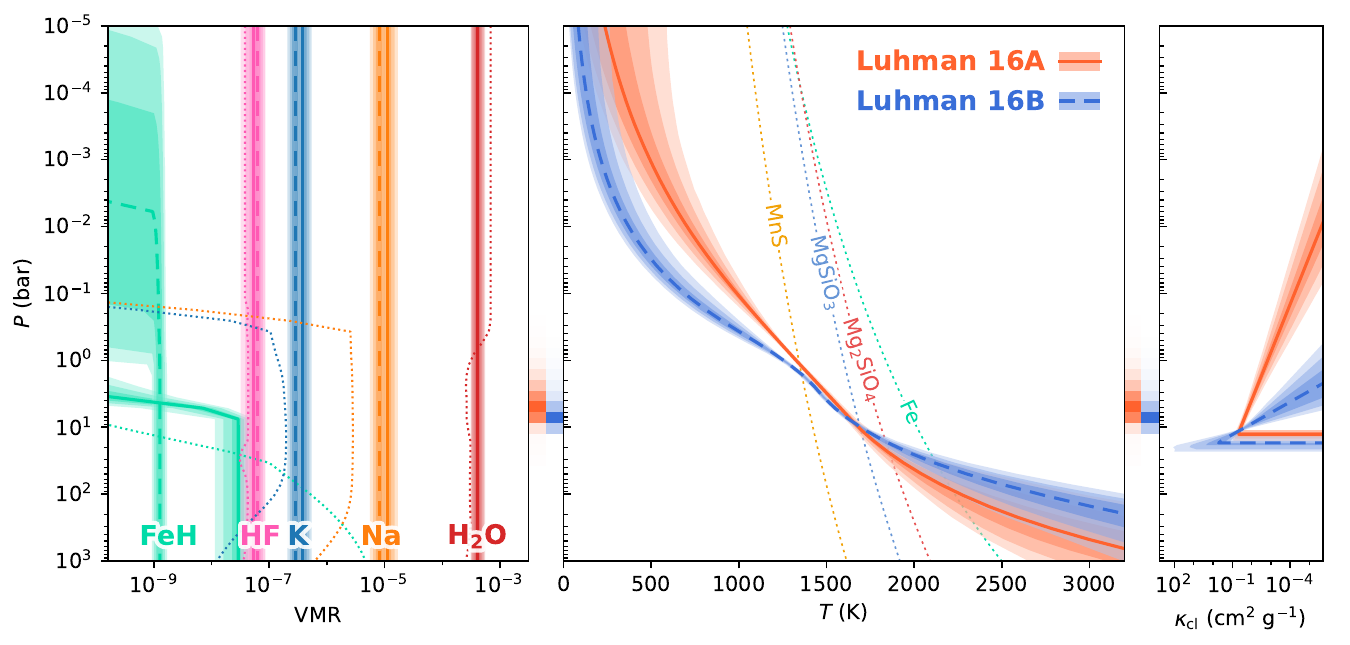}
    \caption{Retrieved vertical profiles of Luhman 16A (solid) and Luhman 16B (dashed). \textit{Left panel}: chemical abundances and the $68$, $95$ and $99.7\%$ confidence envelopes of each species, for both brown dwarfs. The dotted lines show the chemical-equilibrium abundances computed with \texttt{FastChem} \citep{Kitzmann_ea_2024}. With the exception of FeH, the modelled abundances are constant with altitude and thus do not show the drop-offs exhibited by the equilibrium profiles. 
    \textit{Middle panel}: inferred temperature profiles and the condensation curves (dotted) of four cloud species \citep{Visscher_ea_2006,Visscher_ea_2010}. \textit{Right panel}: grey-cloud opacities retrieved as a function of altitude. The orange and blue shading in the panel gaps indicate the Luhman 16AB photospheres as obtained with the integrated emission contribution functions.}
    \label{fig:vertical_profiles}
\end{figure*}

We detect hydrogen-fluoride (HF) and iron-hydride (FeH) in the atmospheres of Luhman 16. The left panels in Fig. \ref{fig:HF_detection} and \ref{fig:FeH_detection} show examples of the contributed absorption from HF and FeH, respectively. The inclusion of either molecule results in a closer match to the observed spectrum of Luhman 16B. To assess the detection significance further, we carry out a cross-correlation analysis between the residuals of a model without species $X$ ($\vec{m}_{i,\mathrm{w/o}\ X}$) and a template which only includes lines from the tested species ($\vec{m}_{i,\mathrm{only}\ X}$). The total cross-correlation function (CCF), at a velocity $\textit{v}$, is then summed over each chip $i$ via
\begin{align}
    \mathrm{CCF}(\textit{v}) &= \sum_i \frac{1}{\tilde{s}_i^2}\left(\vec{d}_i-\tilde{\phi}\vec{m}_{i,\mathrm{w/o}\ X}\right)^T \vec{\Sigma}_{0,i}^{-1}\ \vec{m}_{i,\mathrm{only}\ X}(\textit{v}).
\end{align}
The right panels of Fig. \ref{fig:HF_detection} and \ref{fig:FeH_detection} display this CCF, normalised to a signal-to-noise function using samples outside of the peak ($|\textit{v}|>300\ \mathrm{km\ s^{-1}}$). For Luhman 16B, the cross-correlation yields peak detection significances of $6.7$ and $8.5\sigma$ for HF and FeH, respectively. The detection significances for Luhman 16A are $5.1$ and $10.7\sigma$, respectively, for HF and FeH (see Appendix \ref{app:HF_FeH_A}). The FeH absorption in the J-band spectra of Luhman 16AB was identified previously by \citet{Faherty_ea_2014}. To our knowledge, HF has not been detected in the J-band spectrum of any sub-stellar object. However, recent studies have identified its spectral signature in the K-band \citep{Gonzalez_Picos_ea_2024,Zhang_ea_2024,Mulder_ea_2025} and we provide a discussion in the analysis of the Luhman 16 K-band spectra \citetext{de Regt et al., in prep.}

\subsection{Chemical abundances}
The left panel of Fig. \ref{fig:vertical_profiles} shows the retrieved abundance profiles of Luhman 16A and B as solid and dashed lines, respectively. The two brown dwarfs display remarkably similar abundances, which agree to within $\sim$\,$3\sigma$ for all species except FeH. The dotted lines in the left panel of Fig. \ref{fig:vertical_profiles} indicate the expected chemical-equilibrium abundances, computed using the \texttt{FastChem} code \citep{Kitzmann_ea_2024}. As input, we account for rainout condensation and assume a solar composition (i.e. $\mathrm{[Fe/H]}=0$, $\mathrm{C/O}=0.59$; \citealt{Asplund_ea_2021}). For the most part, the equilibrium abundances of H$_2$O and HF are constant with altitude and in accordance with our retrieved abundances. The equilibrium profiles of Na and K show a dropoff in the gaseous abundance beyond $\sim$\,$0.3\ \mathrm{bar}$ as a result of rainout into sodium- and potassium-bearing condensate species. Our retrieved photospheres are located between $\sim$\,$0.6$--$11\ \mathrm{bar}$, as indicated in the panel gaps, and therefore probe largely constant abundances below this dropoff. We find super-solar alkali abundances where the inferred Na abundance appears elevated by $\sim$\,$0.5\ \mathrm{dex}$, despite the weak absorption seen in Fig. \ref{fig:spectrum}.

\subsection{Clouds and condensation}
The middle panel of Fig. \ref{fig:vertical_profiles} presents the retrieved temperature profiles of Luhman 16A and B as solid orange and dashed blue lines, respectively. In addition, the dotted condensation curves demonstrate the conditions under which the corresponding clouds should form, assuming a solar composition \citep{Visscher_ea_2006,Visscher_ea_2010}. The right panel of Fig. \ref{fig:vertical_profiles} reveals the grey-cloud opacities retrieved for both brown dwarfs. It is interesting to note that the temperature and cloud profiles differ between Luhman 16A and B, whereas the chemical abundances of HF, K, Na, and H$_2$O are in good agreement. This possibly indicates a warming effect by more vertically-extended clouds on Luhman 16A (e.g. \citealt{Morley_ea_2024}). The cloud base pressures ($P_\mathrm{cl,0,A}=13.3^{+1.3}_{-1.2}$, $P_\mathrm{cl,0,B}=18.3^{+1.4}_{-1.1}\ \mathrm{bar}$) coincide with the intersection between the silicate-oxide (MgSiO$_3$, Mg$_2$SiO$_4$) condensation curves and the retrieved PT profiles. These silicate clouds in the Luhman 16AB atmospheres seemingly weaken the J-band spectral features, particularly those formed at high pressures. 

The condensation curve of iron \citep{Visscher_ea_2010} intersects with the Luhman 16AB temperature profiles below the photospheric regions. Iron-bearing species are thus expected to rain out at higher altitudes, thereby decreasing their gaseous abundances. This abundance decrease is also apparent from the equilibrium profile of FeH (dotted turquoise line) in the left panel of Fig. \ref{fig:vertical_profiles}. The FeH abundance of Luhman 16A (solid turquoise line) shows a constraint on a depletion near $\sim$\,$5\ \mathrm{bar}$, although it does not perfectly follow the equilibrium prediction. Conversely, we retrieve FeH at a constant abundance for Luhman 16B (dashed turquoise line). The drop-off pressure for this secondary component is constrained to be above the photospheric altitudes, which leaves the modelled spectrum unaltered. In both brown dwarfs, we find evidence for a higher, non-equilibrium abundance of FeH within the photosphere.

\subsection{Two-column model}
If regions of thinner clouds have formed within the Luhman 16AB photospheres, this could expose deeper, hotter layers of their atmospheres \citep{Faherty_ea_2014}. Our measured elevated FeH abundances might correspond with such clearer patches, similar to the reasoning given for the FeH re-emergence in early T-type brown dwarfs \citep{Burgasser_ea_2002}. We test this hypothesis by introducing a two-column model akin to \citet{Buenzli_ea_2015} and \citet{Vos_ea_2023}. The two columns share most parameters (e.g. PT profile, chemistry, $\log{\textit{g}}$), but have distinct FeH abundances and cloud opacities. The total flux is obtained by summing the separate fluxes, $F_{i}$, while accounting for $\mathcal{CF}$, the surface coverage fraction, 
\begin{align}
    F_\mathrm{tot} &= \mathcal{CF}\cdot F_1 + (1-\mathcal{CF})\cdot F_2.
\end{align}
Thus, an equivalent to the one-column solution can be achieved by setting $\mathcal{CF}=1$ (or $0$) and retrieving the same parameters for column 1 (or column 2). The additional FeH-abundance, cloud-opacity and coverage parameters increase the number of retrieval parameters from 25 to 32 in the two-column model. 

Modelling the Luhman 16A atmosphere as a combination of two columns results in constraints on opaque clouds in both patches (see Table \ref{tab:params}). The smaller patch covers $\sim$\,$6\%$ of the surface and hosts a cloud deck below that of the larger patch ($P_\mathrm{cl,0}=35.9^{+8.8}_{-6.6}$ and $9.8^{+1.1}_{-0.9}\ \mathrm{bar}$). In addition, the minor column finds a constant-with-altitude abundance for FeH ($-7.95^{+0.11}_{-0.08}$), in contrast to the other, FeH-depleted column. In spite of these constraints, a Bayesian evidence comparison reveals that the two-column retrieval is not favoured over the single column ($\ln{B}=-13.3$; $\sim$\,$5.5\sigma$) due to the increased model complexity. 

For Luhman 16B, the Bayesian evidence shows a marginal preference for a two-column solution compared to a single column ($\ln{B}=0.67$; $\sim$\,$1.8\sigma$). However, the constrained parameters are not substantially different from the one-column retrieval. The cloud deck of the one-column solution is replicated by a dominant column that accounts for $\sim$\,$98\%$ of the flux (see Table \ref{tab:params}). The distinct surface gravities constrained in both retrievals hinders a direct comparison of the abundances due to a $\log\textit{g}$-metallicity degeneracy \citep{Zhang_ea_2021,Zhang_ea_2023,de_Regt_ea_2024,Gonzalez_Picos_ea_2024}. Nevertheless, the dominant column seemingly adopts the equivalent FeH abundance ($-9.26^{+0.07}_{-0.08}$) of the one-column retrieval ($-8.90^{+0.05}_{-0.06}$). The minor patch, which covers only $\sim$\,$2\%$ of the modelled surface, hosts no significant cloud opacity and therefore results in a deeper photosphere. At these higher pressures, the retrieval constrains an elevated FeH abundance ($-7.97^{+0.10}_{-0.09}$). 

\section{Discussion} \label{sect:discussion}
\subsection{Surface (in)-homogeneity}
The Luhman 16A spectrum is best described with a single-column model, which points towards a relatively homogeneous surface. This conclusion aligns well with the reduced variability observed for the primary ($\sim$\,$2\%$; \citealt{Biller_ea_2024}) as well as the weak features inferred in Doppler imaging analyses \citep{Crossfield_ea_2014,Chen_ea_2024}. Given its L7.5 spectral type, it is perhaps not surprising that the cloud dissipation expected at the L-T transition has not commenced \citep{Burgasser_ea_2002,Marley_ea_2010}. On the other hand, the marginally preferred two-column model for Luhman 16B ($\sim$\,$1.8\sigma$) constitutes a more ambiguous result. We caution that the constrained coverage fraction, clouds, and FeH abundances are likely biased due to assumptions made to reduce the number of free parameters in the two-column retrieval. For instance, we adopt the same temperature profile for both columns, despite the heating/cooling that should result from the radiative feedback of an opaque cloud \citep{Tan_ea_2019,Tan_ea_2021a,Tan_ea_2021b}. In turn, the altered thermal profile will affect the atmospheric chemistry \citep{Lee_ea_2023,Lee_ea_2024}, which challenges our choice of shared abundances between the two columns. Introducing more parameters to describe separate columns of emission, however, will likely lead to issues of convergence or over-fitting instead of resolving the mentioned ambiguity. Rather, observations over a broader wavelength-range could help to probe the vertical extent of FeH (e.g. in the Y, J, and H bands; \citealt{Faherty_ea_2014}) and the presence of clouds at different altitudes \citep{Vos_ea_2023}. 

\begin{figure*}[h!]
    \centering
    \includegraphics[width=17cm]{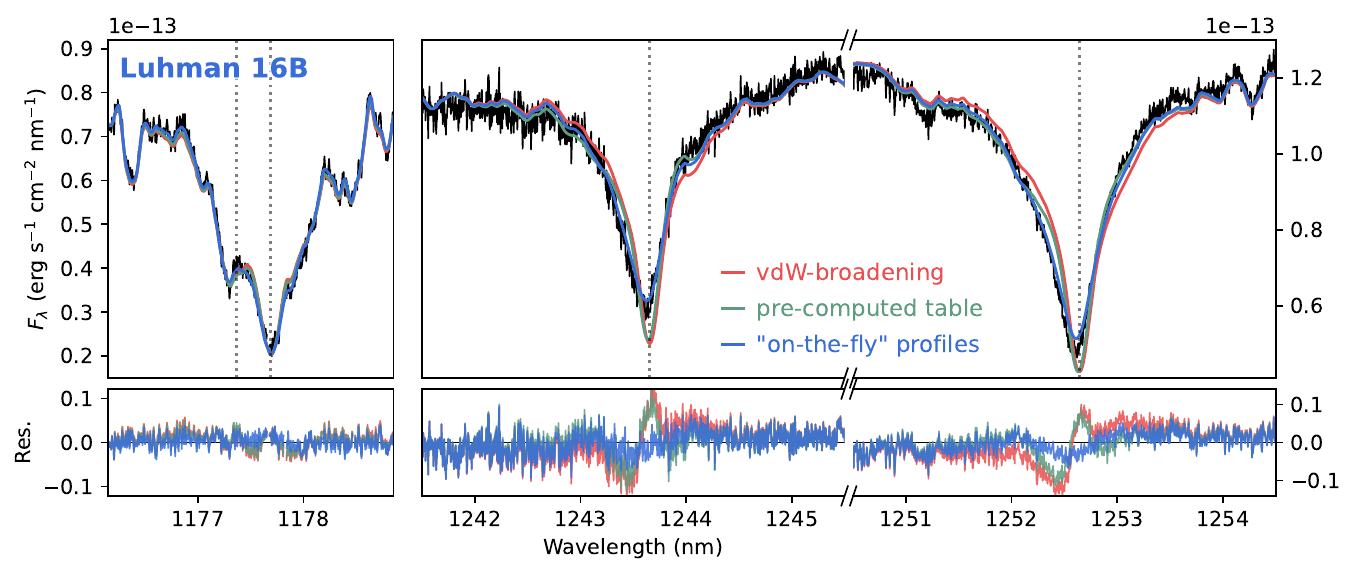}
    \caption{Potassium absorption lines of Luhman 16B, modelled with different approaches. \textit{Left panel}: strong $\mathrm{4p^2P_{3/2}-3d^2D_{3/2}}$ and $\mathrm{4p^2P_{3/2}-3d^2D_{5/2}}$ transitions in the blue end of the J-band. \textit{Right panel}: $\mathrm{4p-5s}$ doublet showing asymmetries compared to the expected transition wavelengths, indicated with dotted lines.}
    \label{fig:K_doublet}
\end{figure*}

\subsection{Alkali line asymmetry} \label{sect:alkali_discussion}
The potassium doublet near $1250\ \mathrm{nm}$ shows an asymmetry in both lines, for both brown dwarfs. Figure \ref{fig:K_doublet} presents an analysis of the potassium absorption lines for Luhman 16B. The red model demonstrates the unsatisfactory fit obtained when only considering van-der-Waals broadening and ignoring line-shifts. It is evident from the right panel of Fig. \ref{fig:K_doublet} that the model profiles, centred on the dotted lines, cannot fit the excess absorption in the blue wing of the lines. The retrieval attempts to alter the photospheric pressures via the surface gravity, temperature profile and cloud deck, resulting in a worsened fit to the bluer J-band doublet (left panel) as well. The model in green utilises pre-computed opacities, calculated on the default $1\ \mathrm{dex}$ pressure-spacing of \texttt{petitRADTRANS} \citep{Molliere_ea_2019}. These opacity cross-sections employ the width- and shift-parameterisation of Sect. \ref{sect:width_shift_param}, but not necessarily at the atmospheric conditions encountered in the Luhman 16 photospheres. This modelling approach results in a better fit to the line wings, but cannot reproduce the line cores adequately. In addition, we find that the model deviates for the other strong alkali lines (left panel) as a result of the interpolation. 

The "on-the-fly" opacity-calculation outlined in Sect. \ref{sect:on_the_fly} results in the best-fitting model, shown in blue in Fig. \ref{fig:K_doublet}. The residuals of the $1175\ \mathrm{nm}$ K doublet are minimised, but even this involved opacity treatment fails to completely reproduce the $1250\ \mathrm{nm}$ profiles. Several factors could contribute to this deviation. First, the line profiles will become non-Lorentzian for perturber densities $\gtrsim$\,$10^{19}\ \mathrm{cm^{-3}}$, akin to the optical resonance lines \citep{Allard_ea_2016,Allard_ea_2024}. Our deepest photospheric layers achieve H$_2$ densities of $n_\mathrm{H_2}\sim3\cdot10^{19}\ \mathrm{cm^{-3}}$, thus challenging the validity of the utilised approach. Furthermore, the assumed $\mathrm{H_2/He}$ abundance ratio affects the total broadening and shift applied to either line. In the future, using high-resolution spectra with sufficient signal-to-noise ratios, one could attempt to infer the H$_2$ and He abundances from the detected line-shifts and -widths. However, since degenerate solutions will likely arise, this is beyond the scope of this study. Additionally, the vertically constant abundance employed for potassium could overestimate the opacity at altitudes where rainout occurs. However, this implies that the current model line core is too deep, contrary to what is seen in Fig. \ref{fig:K_doublet}.

\section{Conclusions} \label{sect:conclusions}
We performed an atmospheric analysis of the nearest brown dwarfs, Luhman 16AB, using high-resolution J-band spectra taken with CRIRES$^+$. The brightness of the two targets provide an unprecedented level of detail for sub-stellar objects. Our \texttt{petitRADTRANS} retrieval models can account for almost all spectral features present in the J-band. Minor deficiencies likely arise from missing absorbers or inaccuracies in the current line lists. These high-quality, high-resolution spectra can form a testbed when updated line lists become available.

For both brown dwarfs, we detect absorption from H$_2$O, K, Na, HF, and FeH. For the first time, this work presents an HF detection in the J-band spectrum of a sub-stellar atmosphere. The retrieval constrains Na (and K) at super-equilibrium abundances in spite of the weak doublet absorption at $\sim$\,$1140\ \mathrm{nm}$ \citep{Faherty_ea_2014,Lodieu_ea_2015}. We find evidence for opaque clouds in the Luhman 16AB photospheres, which are likely made up of silicate-oxide condensates given the constrained cloud deck pressure. The Luhman 16A spectrum is best fit with a homogeneous surface model that includes an FeH-depletion, in line with its reduced variability and L7.5 spectral type \citep{Burgasser_ea_2013,Buenzli_ea_2015,Biller_ea_2024}. We constrain over-abundant FeH in the photosphere of Luhman 16B, which is at odds with the expected condensation into iron clouds. Modelling the Luhman 16B surface as a combination of cloudy and clear patches, however, leads to ambiguous results as this two-column solution is only favoured at $\sim$\,$1.8\sigma$ compared to a homogeneous surface. Observations spanning a wider wavelength range, with JWST for instance, can aid in resolving the altitude-dependent abundance of FeH as well as the vertical extent of cloud decks. 

We detect asymmetric absorption in the $1250\ \mathrm{nm}$ potassium doublet lines ($\mathrm{4p-5s}$) of Luhman 16AB. Employing the unified line shape theory of \citet{Allard_ea_1999}, we parameterise the $PT$-dependent widths and shifts of the Lorentzian profiles. The blue-shifted absorption excess can mostly be reproduced via this careful modelling approach, but the degeneracy with metallicity hinders surface-gravity measurements from our high-resolution spectra alone. Future high-resolution observations of asymmetric atomic lines could help to understand the perturber physics in the sub-stellar and low-mass regime. 

The ESO SupJup Survey observations were mostly taken in the K-band where the $\mathrm{C/O}$-ratio and carbon isotopes can be measured. In that context, the presented analysis provides access to elements that lack absorption features in the K-band for these temperatures, such as Na, K, and Fe. The higher pressures probed in the J-band also permit the investigation of cloud structures. Thus, high-resolution J-band spectra offer valuable insights into the atmospheric properties of Luhman 16AB in particular, and brown dwarfs in general.

\begin{acknowledgements}
We thank the anonymous referee for their constructive comments. 
S.d.R. and I.S. acknowledge funding from NWO grant OCENW.M.21.010. 
Based on observations collected at the European Organisation for Astronomical Research in the Southern Hemisphere under ESO programme(s) 110.23RW.002.
This work used the Dutch national e-infrastructure with the support of the SURF Cooperative using grant no. EINF-4556 and EINF-9460. 
This research has made use of the Astrophysics Data System, funded by NASA under Cooperative Agreement 80NSSC21M00561.
\newline
\textit{Software}: Astropy \citep{Astropy_Collaboration_ea_2022}, corner \citep{Foreman_Mackey_2016}, FastChem \citep{Kitzmann_ea_2024}, Matplotlib \citep{Hunter_2007}, NumPy \citep{Harris_ea_2020}, petitRADTRANS \citep{Molliere_ea_2019}, PyAstronomy \citep{Czesla_ea_2019}, PyMultiNest \citep{Feroz_ea_2009,Buchner_ea_2014}, SciPy \citep{Virtanen_ea_2020}.
\end{acknowledgements}

\bibliographystyle{aa}
\bibliography{references.bib}

\begin{appendix} 
\onecolumn
\section{Contamination correction}\label{app:blend_corr}
\begin{figure*}[h!]
    \centering
    \includegraphics[width=17cm]{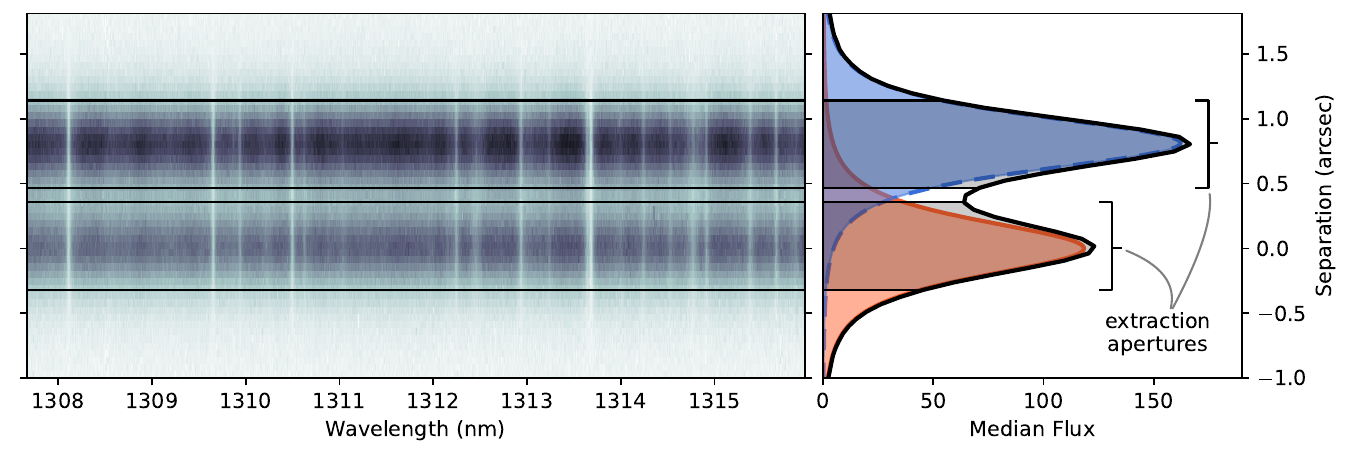}
    \caption{Assessment of the spectral blending in the 23$^\mathrm{rd}$ chip (centred at $\sim$\,$1312\ \mathrm{nm}$). \textit{Left panel}: two-dimensional spectrum, showing emission from Luhman 16B and A at the top and bottom, respectively. The vertical lines show the strong absorption from tellurics. \textit{Right panel}: median spatial profile (black) along with fitted Moffat functions, $\Phi_\mathrm{A}$ (orange) and $\Phi_\mathrm{B}$ (blue). The flux within the 12-pixel wide extraction apertures includes contamination from the unwanted binary component.}
    \label{fig:blend_corr}
\end{figure*}

Figure \ref{fig:blend_corr} demonstrates the spectral contamination between components A and B for a single chip. The left panel displays the two-dimensional spectrum where the slit-tilt and -curvature are rectified in the \texttt{excalibuhr}-reduction. The upper and lower traces are the emission of Luhman 16B and A, respectively. The spatial profile, median-combined over all wavelength channels in the chip, is shown in black in the right panel of Fig. \ref{fig:blend_corr}. While the peak emissions can be distinguished at a separation of $\sim$\,$0.81"$ (as predicted by \citealt{Bedin_ea_2024}), there exists $\sim$\,$6$--$11\%$ contamination within the respective 12-pixel wide extraction apertures. 

To correct the contamination, we fit two Moffat functions\footnote{\url{https://docs.astropy.org/en/latest/api/astropy.modeling.functional_models.Moffat1D.html}} to the observed spatial profile. These fitted profiles, $\Phi_\mathrm{A}$ and $\Phi_\mathrm{B}$, are shown in orange and blue, respectively, in the right panel of Fig. \ref{fig:blend_corr}. The profiles are subsequently integrated over the extraction apertures to assess the relative contributions. The measured flux in aperture A is then the weighted sum of the uncontaminated total fluxes of components A and B. That is, 
\begin{align}
    F_\mathrm{in\ aper.\ A} &= \left(\frac{\int_\mathrm{aper.\ A}\Phi_\mathrm{A}}{\int\Phi_\mathrm{A}}\right)\cdot F_\mathrm{tot,A} + \left(\frac{\int_\mathrm{aper.\ A}\Phi_\mathrm{B}}{\int\Phi_\mathrm{B}}\right)\cdot F_\mathrm{tot,B}, 
\end{align}
where the integrals over aperture A are normalised by integrals over the complete profiles. Similarly, the measured flux in aperture B is expressed as
\begin{align}
    F_\mathrm{in\ aper.\ B} &= \left(\frac{\int_\mathrm{aper.\ B}\Phi_\mathrm{A}}{\int\Phi_\mathrm{A}}\right)\cdot F_\mathrm{tot,A} + \left(\frac{\int_\mathrm{aper.\ B}\Phi_\mathrm{B}}{\int\Phi_\mathrm{B}}\right)\cdot F_\mathrm{tot,B}. 
\end{align}
Through substitution, we obtain definitions for the blending-corrected fluxes of Luhman 16A and B, 
\begin{align}
    F_\mathrm{A} &= F_\mathrm{in\ aper.\ A} - \left(\frac{\int_\mathrm{aper.\ A}\Phi_\mathrm{B}}{\int_\mathrm{aper.\ B}\Phi_\mathrm{B}}\right)\cdot F_\mathrm{in\ aper.\ B}, \\
    F_\mathrm{B} &= F_\mathrm{in\ aper.\ B} - \left(\frac{\int_\mathrm{aper.\ B}\Phi_\mathrm{A}}{\int_\mathrm{aper.\ A}\Phi_\mathrm{A}}\right)\cdot F_\mathrm{in\ aper.\ A}.
\end{align}
We note that $F_\mathrm{A}\neq F_\mathrm{tot,A}$ (and $F_\mathrm{B}\neq F_\mathrm{tot,B}$) because a scaling factor is omitted. The absolute scale of $F_\mathrm{A}$ (and $F_\mathrm{B}$), however, is not important because the spectrum is flux-calibrated after the outlined contamination correction (see Sect. \ref{sect:obs_and_red}). The seeing-limited point-spread function becomes narrower at longer wavelengths and we therefore perform the correction separately for each chip. 
\newpage

\section{Retrieved parameters}
\begin{table*}[h!]
    \centering
    \caption{Retrieved parameters and their uncertainties.}
    \label{tab:params}
    \begin{tabular}{lll|rr|rr}
        \hline
        \hline
        \textbf{Parameter} & \textbf{Description} & \textbf{Prior} & \multicolumn{2}{c|}{\textbf{Luhman 16A}} & \multicolumn{2}{c}{\textbf{Luhman 16B}} \\
         & & & \textbf{1-column} & \textbf{2-column} & \textbf{1-column} & \textbf{2-column} \\
        \hline
        $\log\mathrm{H_2O}$ & H$_2$O abundance & $\mathcal{U}(-14.0,-2.0)$ & $-3.38^{+0.06}_{-0.05}$ & $-3.42^{+0.04}_{-0.05}$ & $-3.38^{+0.04}_{-0.05}$ & $-3.48^{+0.03}_{-0.03}$ \\
        $\log\mathrm{Na}$   & Na abundance     & $\mathcal{U}(-14.0,-2.0)$ & $-4.95^{+0.07}_{-0.06}$ & $-4.97^{+0.05}_{-0.06}$ & $-5.09^{+0.05}_{-0.06}$ & $-5.15^{+0.05}_{-0.04}$ \\
        $\log\mathrm{K}$    & K abundance      & $\mathcal{U}(-14.0,-2.0)$ & $-6.42^{+0.07}_{-0.06}$ & $-6.44^{+0.05}_{-0.06}$ & $-6.55^{+0.05}_{-0.05}$ & $-6.61^{+0.04}_{-0.03}$ \\
        $\log\mathrm{HF}$   & HF abundance     & $\mathcal{U}(-14.0,-2.0)$ & $-7.28^{+0.07}_{-0.07}$ & $-7.32^{+0.06}_{-0.07}$ & $-7.21^{+0.05}_{-0.06}$ & $-7.31^{+0.04}_{-0.04}$ \\
        $\log\mathrm{(FeH)_0}$ & FeH abundance at $P\geq P_{\mathrm{FeH},0}$ & $\mathcal{U}(-14.0,-2.0)$ & 
            $-7.54^{+0.10}_{-0.11}$ & \begin{tabular}{@{}r@{}}\textcolor{c1}{$-11.07^{+1.71}_{-1.49}$}\\\textcolor{c2}{$-7.95^{+0.11}_{-0.08}$}\end{tabular} & 
            $-8.90^{+0.05}_{-0.06}$ & \begin{tabular}{@{}r@{}}\textcolor{c1}{$-7.97^{+0.10}_{-0.09}$}\\\textcolor{c2}{$-9.26^{+0.07}_{-0.08}$}\end{tabular} \\
        $\log P_{\mathrm{FeH},0}$ & FeH drop-off pressure & $\mathcal{U}(-5.0,3.0)$ & 
            $+0.78^{+0.04}_{-0.03}$ & \begin{tabular}{@{}r@{}}\textcolor{c1}{$-1.4^{+2.1}_{-1.8}$}\\\textcolor{c2}{$-0.9^{+1.4}_{-2.0}$}\end{tabular} & 
            $-2.2^{+1.5}_{-1.5}$  & \begin{tabular}{@{}r@{}}\textcolor{c1}{$-2.1^{+1.7}_{-1.5}$}\\\textcolor{c2}{$-2.5^{+1.2}_{-1.3}$}\end{tabular} \\
        $\alpha_\mathrm{FeH}$ & FeH drop-off power & $\mathcal{U}(0.0,20.0)$ & 
            $9.5^{+4.4}_{-3.2}$ & \begin{tabular}{@{}r@{}}\textcolor{c1}{$9.2^{+5.1}_{-4.9}$}\\\textcolor{c2}{$9.2^{+5.1}_{-4.5}$}\end{tabular} & 
            $9.6^{+5.4}_{-5.2}$ & \begin{tabular}{@{}r@{}}\textcolor{c1}{$9.8^{+5.1}_{-4.9}$}\\\textcolor{c2}{$9.4^{+5.2}_{-5.0}$}\end{tabular} \\
        \hline
        $\log\textit{g}$            & Surface gravity               & \begin{tabular}{@{}r@{}}$\mathcal{G}_\mathrm{A}(4.96,0.09)$\\$\mathcal{G}_\mathrm{B}(4.89,0.09)$\end{tabular} & $4.95^{+0.06}_{-0.05}$ & $4.95^{+0.04}_{-0.05}$ & $4.89^{+0.04}_{-0.05}$ & $4.80^{+0.03}_{-0.03}$ \\
        $\textit{v}\sin i$          & Projected rotational velocity & $\mathcal{U}(10.0,30.0)$ & $14.93^{+0.14}_{-0.12}$ & $14.87^{+0.11}_{-0.10}$ & $24.48^{+0.04}_{-0.03}$ & $24.47^{+0.03}_{-0.03}$ \\
        $\varepsilon_\mathrm{limb}$ & Limb-darkening coefficient    & $\mathcal{U}(0.0,1.0)$   & $0.46^{+0.06}_{-0.06}$ & $0.44^{+0.05}_{-0.04}$ & $0.01^{+0.01}_{-0.01}$ & $0.01^{+0.01}_{-0.01}$ \\
        $\textit{v}_\mathrm{rad}$   & Radial velocity               & $\mathcal{U}(10.0,30.0)$ & $17.27^{+0.03}_{-0.03}$ & $17.27^{+0.02}_{-0.02}$ & $20.00^{+0.04}_{-0.03}$ & $20.01^{+0.03}_{-0.03}$ \\
        $\mathcal{CF}$              & Surface coverage fraction     & $\mathcal{U}(0.0,1.0)$   & & \begin{tabular}{@{}r@{}}\textcolor{c1}{$0.942^{+0.006}_{-0.006}$}\\\textcolor{c2}{$0.058^{+0.006}_{-0.006}$}\end{tabular} & & \begin{tabular}{@{}r@{}}\textcolor{c1}{$0.024^{+0.006}_{-0.004}$}\\\textcolor{c2}{$0.976^{+0.004}_{-0.006}$}\end{tabular} \\
        \hline
        $\nabla_1$ & Temperature gradient at $P_1$ & $\mathcal{U}(0.10,0.34)$ & $0.22^{+0.07}_{-0.07}$ & $0.22^{+0.06}_{-0.06}$ & $0.22^{+0.06}_{-0.06}$ & $0.22^{+0.06}_{-0.06}$ \\
        $\nabla_2$ & Temperature gradient at $P_2$ & $\mathcal{U}(0.10,0.34)$ & $0.17^{+0.06}_{-0.04}$ & $0.15^{+0.03}_{-0.02}$ & $0.25^{+0.05}_{-0.05}$ & $0.22^{+0.04}_{-0.04}$ \\
        $\nabla_3$ & Temperature gradient at $P_3$ & $\mathcal{U}(0.05,0.34)$ & $0.090^{+0.007}_{-0.007}$ & $0.094^{+0.004}_{-0.004}$ & $0.083^{+0.010}_{-0.009}$ & $0.093^{+0.010}_{-0.009}$ \\
        $\nabla_4$ & Temperature gradient at $P_4$ & $\mathcal{U}(0.0,0.34)$  & $0.15^{+0.02}_{-0.02}$ & $0.16^{+0.03}_{-0.02}$ & $0.30^{+0.02}_{-0.03}$ & $0.29^{+0.03}_{-0.03}$ \\
        $\nabla_5$ & Temperature gradient at $P_5$ & $\mathcal{U}(0.0,0.34)$  & $0.16^{+0.10}_{-0.09}$ & $0.19^{+0.08}_{-0.09}$ & $0.18^{+0.09}_{-0.09}$ & $0.17^{+0.09}_{-0.09}$ \\
        $\log P_3$ & Pressure of central knot      & $\mathcal{U}(-1.0,1.0)$  & $+0.90^{+0.06}_{-0.08}$ & $+0.53^{+0.12}_{-0.12}$ & $+0.43^{+0.07}_{-0.08}$ & $+0.38^{+0.08}_{-0.08}$ \\
        $T_3$      & Temperature at central knot   & $\mathcal{U}(1200,2200)$ & $1644^{+23}_{-28}$ & $1505^{+42}_{-43}$ & $1444^{+27}_{-28}$ & $1418^{+29}_{-31}$ \\
        $\Delta\log P_{23}$ & Separation between $P_2$ and $P_3$ & $\mathcal{U}(-2.0,-0.5)$ & $-1.4^{+0.4}_{-0.3}$ & $-1.3^{+0.3}_{-0.3}$ & $-1.6^{+0.3}_{-0.2}$ & $-1.6^{+0.3}_{-0.2}$ \\
        $\Delta\log P_{34}$ & Separation between $P_3$ and $P_4$ & $\mathcal{U}(0.5,3.0)$   & $+2.3^{+0.4}_{-0.5}$ & $+2.2^{+0.4}_{-0.5}$ & $+1.0^{+0.4}_{-0.3}$ & $+1.1^{+0.4}_{-0.3}$ \\
        \hline
        $\log\kappa_{\mathrm{cl},0}$ & Cloud-base opacity & $\mathcal{U}(-10.0,3.0)$ & 
            $-1.27^{+0.09}_{-0.09}$ & \begin{tabular}{@{}r@{}}\textcolor{c1}{$+1.1^{+0.8}_{-0.8}$}\\\textcolor{c2}{$+1.2^{+0.9}_{-1.0}$}\end{tabular} & 
            $-0.05^{+0.72}_{-0.41}$ & \begin{tabular}{@{}r@{}}\textcolor{c1}{$-5.5^{+2.7}_{-2.4}$}\\\textcolor{c2}{$+0.2^{+0.7}_{-0.5}$}\end{tabular} \\
        $\log P_{\mathrm{cl},0}$ & Cloud-base pressure & $\mathcal{U}(0.0,2.5)$ & 
            $+1.12^{+0.04}_{-0.04}$ & \begin{tabular}{@{}r@{}}\textcolor{c1}{$+0.99^{+0.05}_{-0.04}$}\\\textcolor{c2}{$+1.56^{+0.10}_{-0.09}$}\end{tabular} & 
            $+1.26^{+0.03}_{-0.03}$ & \begin{tabular}{@{}r@{}}\textcolor{c1}{$+0.97^{+0.82}_{-0.81}$}\\\textcolor{c2}{$+1.28^{+0.03}_{-0.03}$}\end{tabular} \\
        $f_\mathrm{sed}$ & Cloud-opacity decay & $\mathcal{U}(1.0,20.0)$ & 
            $1.4^{+0.2}_{-0.2}$ & \begin{tabular}{@{}r@{}}\textcolor{c1}{$10.3^{+2.3}_{-2.3}$}\\\textcolor{c2}{$12.3^{+3.2}_{-3.1}$}\end{tabular} & 
            $6.1^{+2.7}_{-1.4}$ & \begin{tabular}{@{}r@{}}\textcolor{c1}{$10.6^{+4.9}_{-5.0}$}\\\textcolor{c2}{$6.5^{+2.0}_{-1.4}$}\end{tabular} \\
        \hline
        $\log a$    & GP amplitude    & $\mathcal{U}(-0.7,0.3)$  & $-0.219^{+0.006}_{-0.006}$ & $-0.217^{+0.005}_{-0.005}$ & $-0.156^{+0.005}_{-0.005}$ & $-0.158^{+0.005}_{-0.005}$ \\
        $\log \ell$ & GP length-scale & $\mathcal{U}(-3.0,-1.0)$ & $-1.376^{+0.012}_{-0.012}$ & $-1.377^{+0.010}_{-0.009}$ & $-1.285^{+0.006}_{-0.005}$ & $-1.285^{+0.006}_{-0.005}$ \\
        \hline
         & & \multicolumn{1}{r|}{$\ln\ B$} & & $-13.3$ & & $+0.67$ \\
    \end{tabular}
    \tablefoot{All priors are uniform (indicated with $\mathcal{U}(\mathrm{min},\mathrm{max})$), except for the surface gravity, which is retrieved using a Gaussian prior ($\mathcal{G}(\mu,\sigma)$) and different between Luhman 16A and B. In the case of the FeH abundance and cloud-opacity parameters, the two-column solution shows the two values corresponding to each patch in black and grey. The surface coverage fraction is also calculated for the second column (as $1-\mathcal{CF}$), but it is only retrieved as a single parameter.}
\end{table*}
\newpage

\begin{figure*}[h!]
    \centering
    \includegraphics[width=17cm]{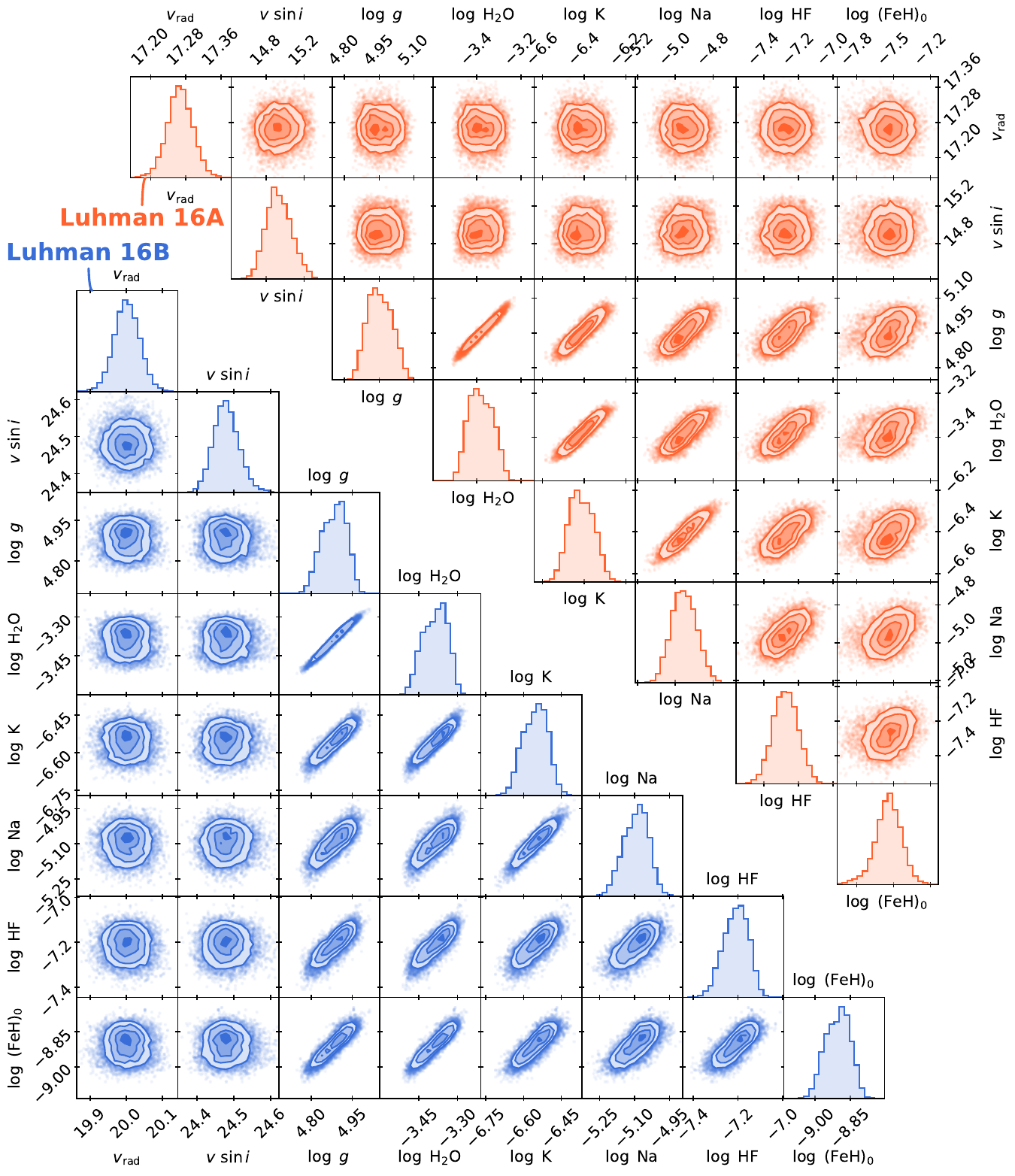}
    \caption{Posterior distributions of a selection of parameters for the Luhman 16AB one-column retrievals. For both objects, we find a strong correlation between the surface gravity, $\log{\textit{g}}$, and the abundances. However, the retrieved surface gravities do not deviate from the imposed Gaussian priors which implies that the constrained abundances are accurate. The radial velocity, $\textit{v}_\mathrm{rad}$, shows no correlation and could therefore help to constrain the Luhman 16AB orbits further.}
    \label{fig:corner}
\end{figure*}

\newpage
\section{Best-fitting spectra} \label{app:best_fitting_spectra}
\begin{figure*}[h!]
    \centering
    \includegraphics[width=17cm]{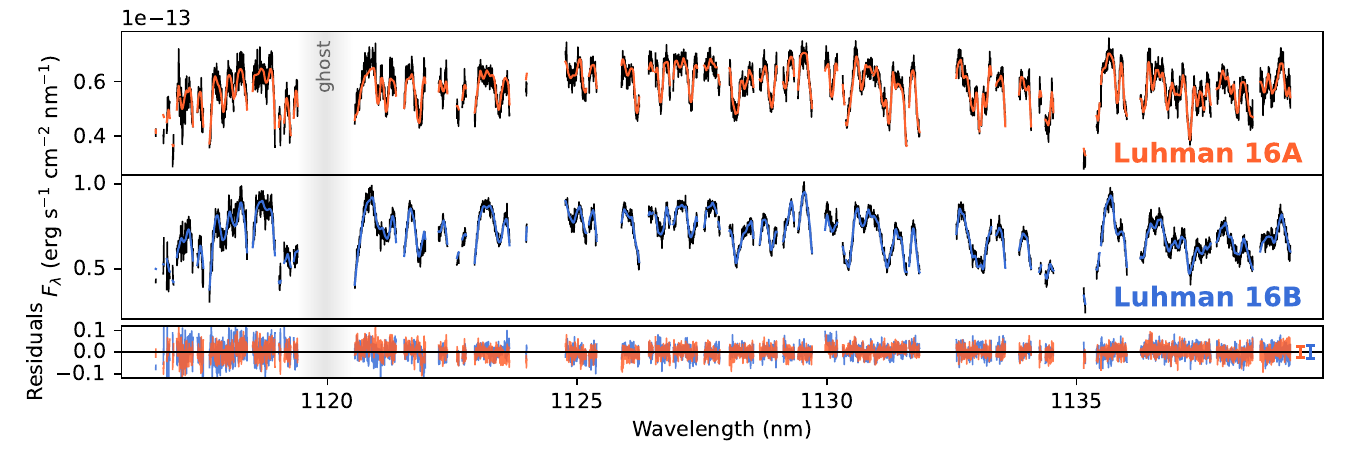}
    \includegraphics[width=17cm]{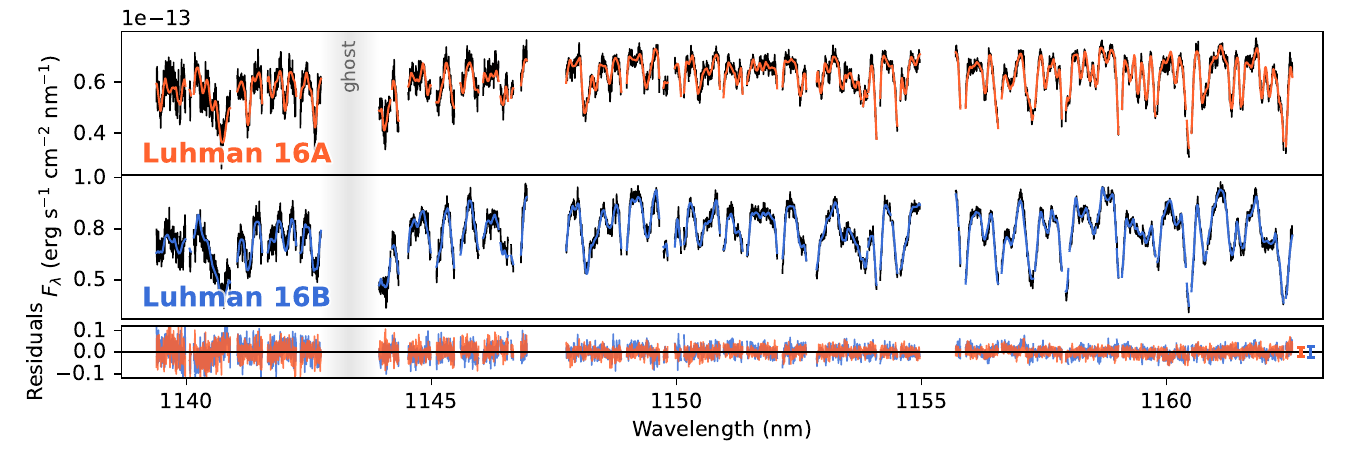}
    \includegraphics[width=17cm]{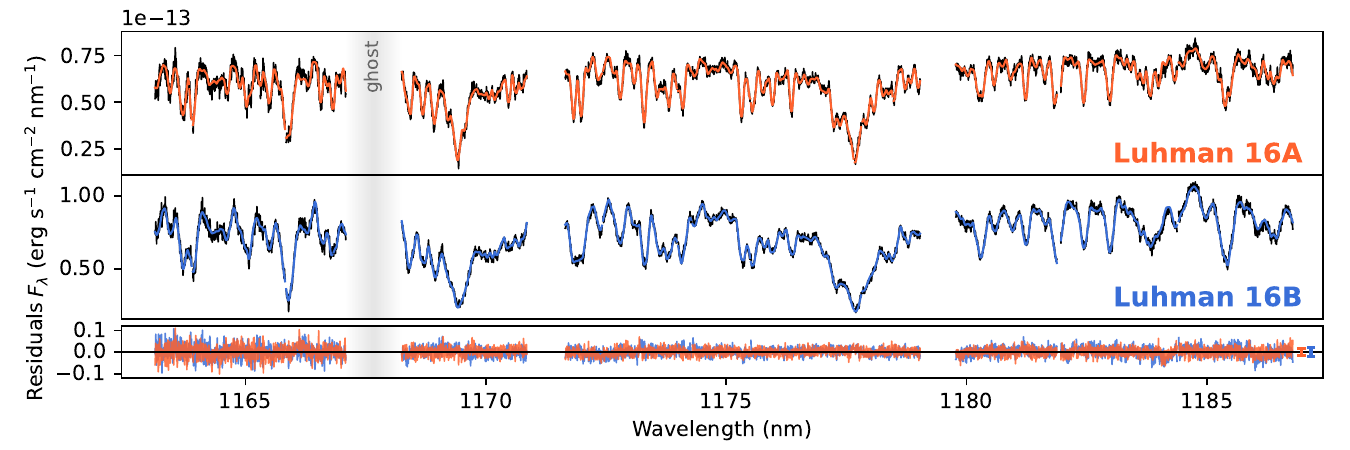}
    \includegraphics[width=17cm]{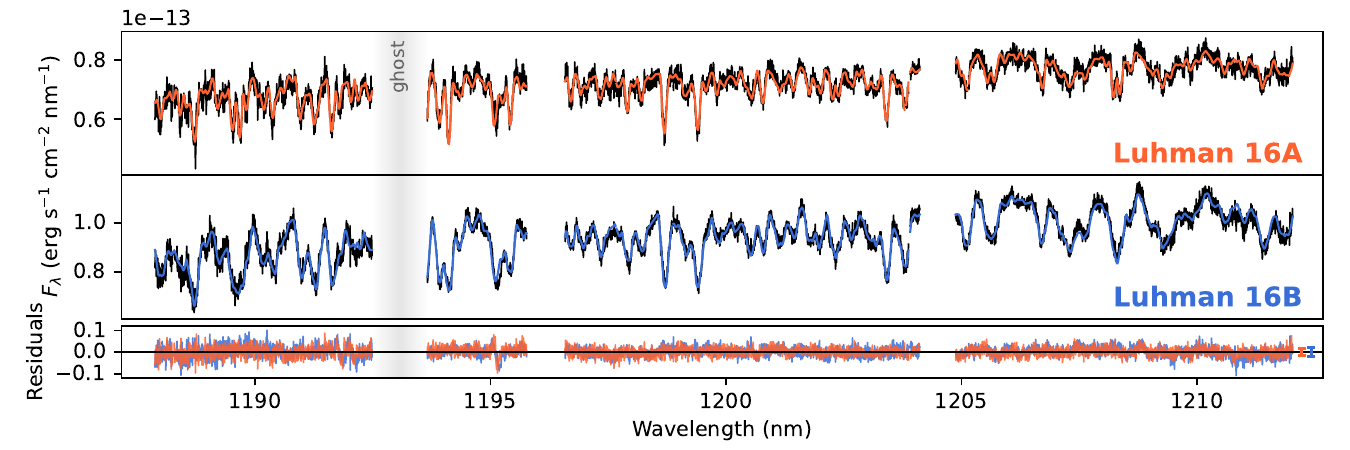}
    \caption{Same as Fig. \ref{fig:bestfit_spectrum}, but showing all spectral orders.}
\end{figure*}

\newpage
\begin{figure*}[h!]
    \centering
    \includegraphics[width=17cm]{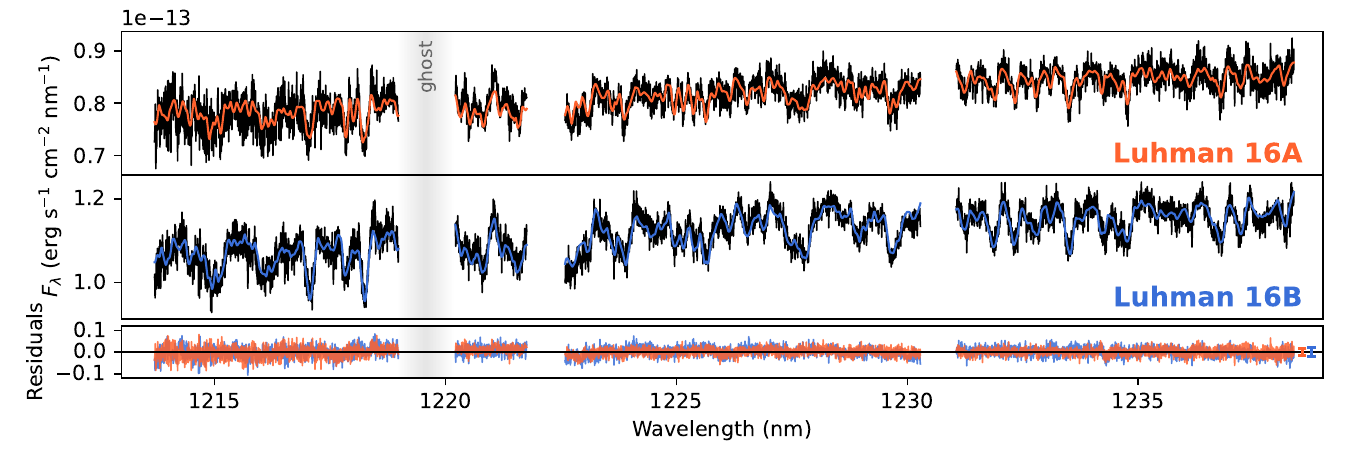}
    \includegraphics[width=17cm]{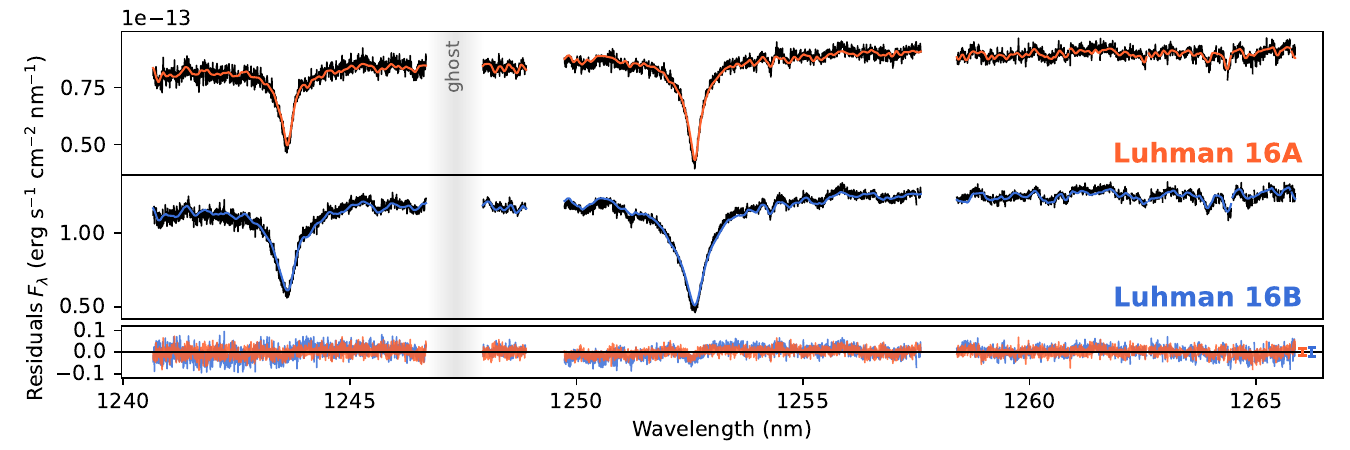}
    \includegraphics[width=17cm]{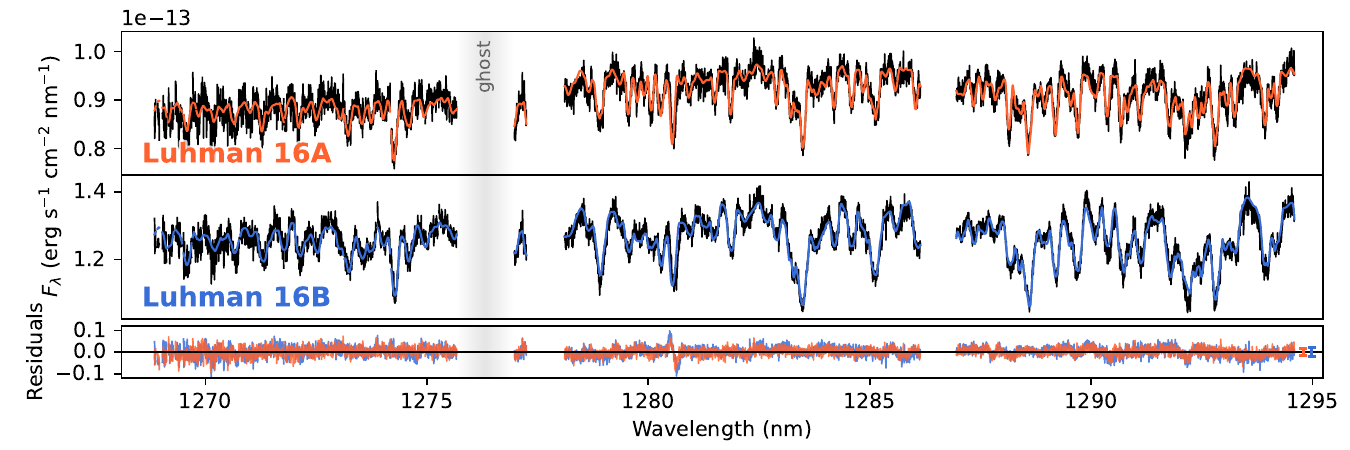}
    \includegraphics[width=17cm]{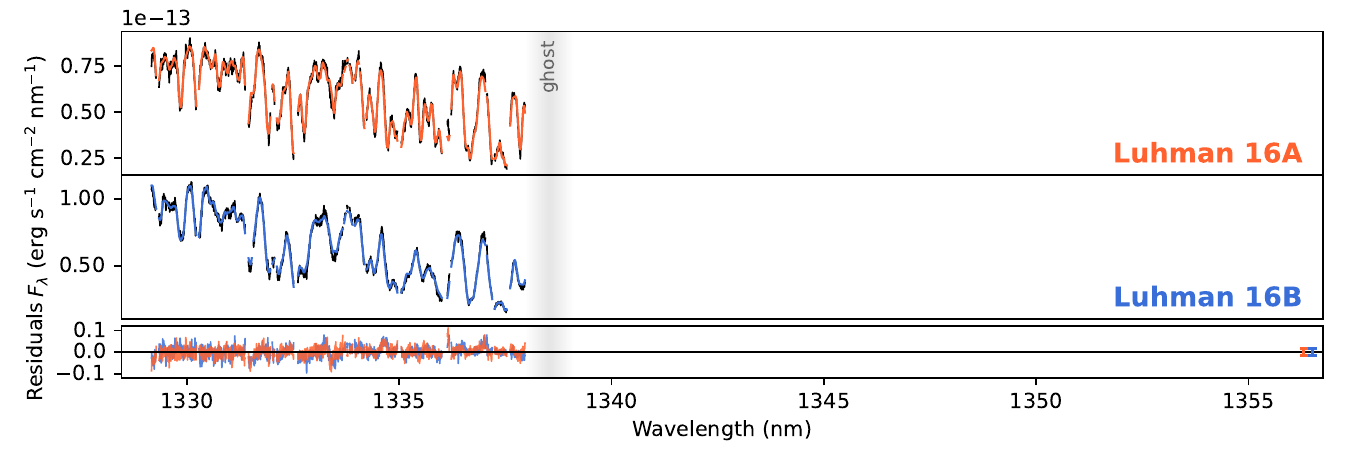}
    \caption{Continued. The missing spectral order is shown in Fig. \ref{fig:bestfit_spectrum}.}
\end{figure*}

\newpage
\section{HF and FeH detection on Luhman 16A} \label{app:HF_FeH_A}
\begin{figure*}[h!]
    \centering
    \includegraphics[width=17cm]{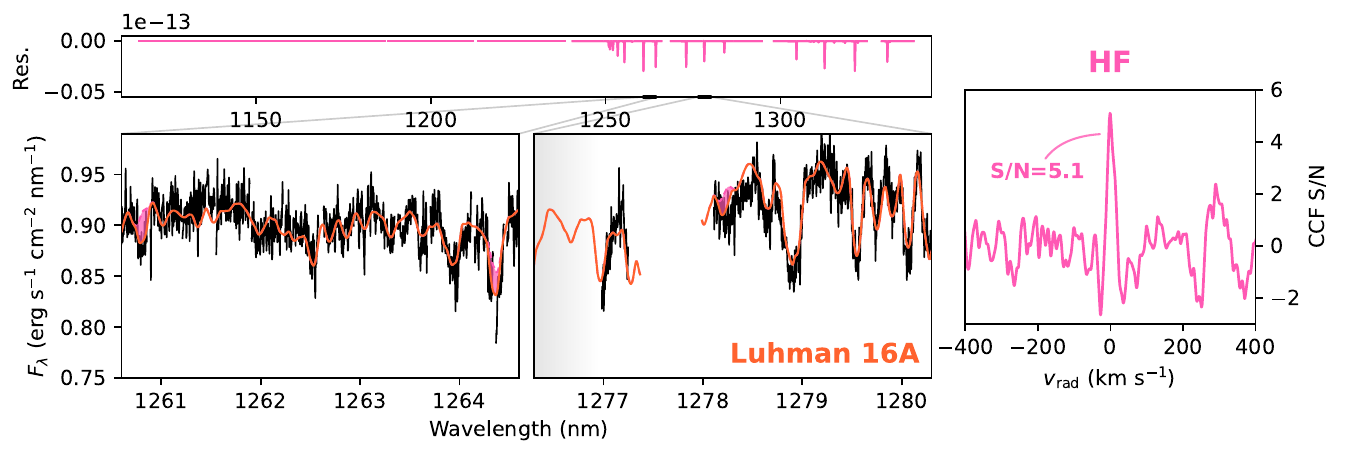}
    \caption{Detection analysis of hydrogen-fluoride (HF) in the Luhman 16A spectrum, same as Fig. \ref{fig:HF_detection}.}
\end{figure*}

\begin{figure*}[h!]
    \centering
    \includegraphics[width=17cm]{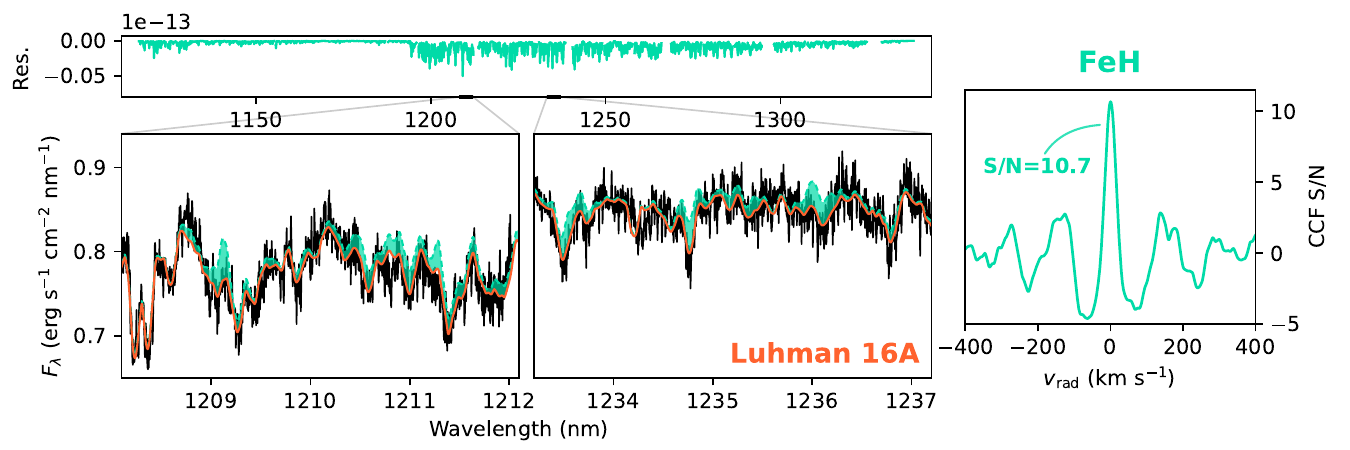}
    \caption{Same as Fig. \ref{fig:FeH_detection}, but for Luhman 16A.}
\end{figure*}

\end{appendix}

\end{document}